\documentclass[aps,a4paper,superscriptaddress,nofootinbib,twocolumn,journal,twoside]{revtex4-1}
\usepackage{amsmath,amssymb,graphicx,multirow,dcolumn,bm,latexsym,soul,nicefrac,acronym}
\usepackage{appendix}
\usepackage[colorlinks,linkcolor=blue,citecolor=blue,urlcolor=blue ]{hyperref}
\usepackage{float}
\usepackage{soul}
\usepackage{verbatim}
\usepackage{color}
\usepackage{acronym}
\usepackage{subfigure}
\usepackage{appendix}
\usepackage{longtable}
\usepackage{footnote}
\makeatletter
\newcommand{\be}{\begin{equation}}
\newcommand{\ee}{\end{equation}}
\newcommand{\bea}{\begin{eqnarray}}
\newcommand{\eea}{\end{eqnarray}}
\newcommand{\nn}{\nonumber}

\newcommand{\wtd}{\widetilde}
\newcommand{\eq}[1]{Eq.~(\ref{#1})}
\newcommand{\eqs}[1]{Eqs.~(\ref{#1})}
\newcommand{\fig}[1]{Fig.~\ref{#1}}
\newcommand{\figs}[1]{Figs.~\ref{#1}}
\newcommand{\tab}[1]{Table.~\ref{#1}}

\newcommand{\TRC}{MOE Key Laboratory of TianQin Mission, TianQin Research Center for Gravitational Physics $\&$  School of Physics and Astronomy, Frontiers Science Center for TianQin, CNSA Research Center for Gravitational Waves, Sun Yat-sen University (Zhuhai Campus), Zhuhai 519082, China}
\newcommand{\HUST}{MOE Key Laboratory of Fundamental Physical Quantities Measurements,
Hubei Key Laboratory of Gravitation and Quantum Physics, School of Physics,
Huazhong University of Science and Technology, Wuhan 430074, China}

\begin{document}
\title{Science with the TianQin Observatory:\\Preliminary Results on stochastic gravitational-wave background}
\author{Zheng-Cheng Liang}
\email{zhchliang@hust.edu.cn}
\affiliation{\HUST}
\affiliation{\TRC}
\author{Yi-Ming Hu}
\email{huyiming@sysu.edu.cn}
\affiliation{\TRC}
\author{Yun Jiang}
\email{jiangyun5@sysu.edu.cn}
\affiliation{\TRC}
\affiliation{Max-Planck-Institut f\"ur Kernphysik, Saupfercheckweg 1, 69117 Heidelberg, Germany}
\author{Jun Cheng}
\email{chengj79@mail2.sysu.edu.cn}
\affiliation{\TRC}
\author{Jian-dong Zhang}
\email{zhangjd9@sysu.edu.cn}
\affiliation{\TRC}
\author{Jianwei Mei}
\email{meijw@sysu.edu.cn}
\affiliation{\TRC}

%\input{git_tag.tex}
%\date{\commitDATE; \commitIDshort-\commitSTATUS}

\date{\today}

\begin{abstract}
In this work, we study the prospect of detecting the stochastic gravitational-wave background with the TianQin Observatory.
We consider sources of both astrophysical-origin and cosmological-origin, including stellar-mass binary black holes, binary neutron stars, Galactic white dwarves, inflation, first-order phase transitions, and cosmic defects.
For the detector configurations, we consider TianQin, TianQin I+II, and TianQin + LISA. 
We study the detectability of stochastic gravitational-wave backgrounds with both the cross correlation and null channel methods, and present the corresponding power-law integrated sensitivity curves. 
We introduce the definition of the ``joint foreground'' with a network of detectors. 
With the joint foreground, the number of resolved double white dwarves in the Galaxy will be increased by 5$-$22\% compared with a simple combination of individual detectors.
The astrophysical background is expected to be detectable with a signal-to-noise ratio of 100 after 5 years of operation and dominated by the extragalactic double white dwarves. 
On the other hand, due to the uncertain nature of underlying models, we can
only estimate the detection capability of the cosmological background for
specific cases.
\end{abstract}



%%%%%%%%%%%%%%%%%%%%%%%%%%%%%%%%%%%%%%%%%%%%%%%%%%%%%%%%%%%%%%%%
\maketitle
\acrodef{SGWB}{stochastic GW background}
\acrodef{GW}{gravitational-wave}
\acrodef{CBC}{compact binary coalescence}
\acrodef{MBHB}{massive black hole binary}
\acrodef{SBBH}{stellar-mass binary black hole}
\acrodef{EMRI}{extreme-mass-ratio inspiral}
\acrodef{DWD}{double white dwarf}
\acrodef{EDWD}{extragalactic double white dwarf}
\acrodef{BH}{black hole}
\acrodef{NS}{neutron star}
\acrodef{BNS}{binary neutron star}
\acrodef{LIGO}{Laser Interferometer Gravitational-Wave Observatory}
\acrodef{TQ}{TianQin}
\acrodef{LISA}{Laser Interferometer Space Antenna}
\acrodef{KAGRA}{Kamioka Gravitational Wave Detector}
\acrodef{ET}{Einstein telescope}
\acrodef{DECIGO}{DECi-hertz Interferometer GravitationalWave Observatory}
\acrodef{CE}{Cosmic Explorer}
\acrodef{NANOGrav}{The North American Nanohertz Observatory for Gravitational Waves}
\acrodef{ORF}{overlap reduction function}
\acrodef{ASD}{amplitude spectrum density}
\acrodef{PSD}{power spectrum density}
\acrodef{SNR}{signal-to-noise ratio}
\acrodef{TDI}{time delay interferometer}
\acrodef{PI}{power-law integrated}
\acrodef{PBH}{primordial black hole}
\acrodef{SSB}{solar system baryo}
\acrodef{PT}{phase transition}
\acrodef{SM}{Standard Model}
\acrodef{EWPT}{electroweak phase transition}
\acrodef{PTA}{Pulsar Timing Arrays}
\acrodef{RD}{radiation-dominated}
\acrodef{MD}{matter-dominated}
\acrodef{NG}{Nambu-Goto}
%%%%%%%%%%%%%%%%%%%%%%%%%%%%%%%%%%%%%%%%%%%%%%%%%%%%%%%%%%%%%%%%
%%%% ±êÌâÒ³œáÊø %%%%%%%%%%%%%%%%%%%%%%%%%%%%%%%%%%%%%%%%%%%%%%%%
%%%%%%%%%%%%%%%%%%%%%%%%%%%%%%%%%%%%%%%%%%%%%%%%%%%%%%%%%%%%%%%%
%%%% µÚÒ»œÚ¿ªÊŒ %%%%%%%%%%%%%%%%%%%%%%%%%%%%%%%%%%%%%%%%%%%%%%%%
%%%%%%%%%%%%%%%%%%%%%%%%%%%%%%%%%%%%%%%%%%%%%%%%%%%%%%%%%%%%%%%%

\section{Introduction}
%==========The definition of \ac{SGWB}================

Gravitational-wave (GW) emission from a large number of independent and unresolvable sources cannot be detected individually by detectors. 
Nevertheless, their incoherent superposition would form a \ac{SGWB}, which could be detected as a bulk~\cite{Maggiore:1999vm,Christensen:2018iqi,Romano:2019yrj}. 
Many different mechanisms can contribute to the \ac{SGWB}, by which the \ac{SGWB} can be roughly divided into two categories, {i.e.,} of either astrophysical-origin and cosmological-origin~\cite{deAraujo:2000gw,Martinovic:2020hru}. 

%==========The origin of \ac{SGWB}: a.astrophysics; b. cosmology. The significance of hunting for \ac{SGWB}================
The astrophysical \ac{SGWB} mainly contains the GW emission from plenty of compact binaries~\cite{Audley:2017drz}.
For space-borne GW detectors, such sources could be \acp{DWD}~\cite{Korol:2017qcx,Huang:2020rjf}, \acp{MBHB}~\cite{Wang:2019}, \acp{SBBH}~\cite{Liu:2020eko}, and \acp{EMRI}~\cite{Fan:2020zhy,Zi:2021pdp}. 
On the other hand, the cosmological \ac{SGWB} for space-borne GW detectors contains the GWs from physical processes linked to the early Universe, like inflation~\cite{Guth:1982ec}, first-order \acp{PT}~\cite{Hogan:1984hx}, and cosmic defects~\cite{Kibble:1976sj}. 
A detection of the \ac{SGWB} would have important implications in either astrophysics and cosmology. 
When exploring the \ac{SGWB} of extragalactic compact binaries, one can limit the event rate, mass distribution, and formation mechanisms~\cite{Mazumder:2014fja,Callister:2016ewt,Maselli:2016ekw}. 
As for the \ac{SGWB} of \acp{DWD} in the Galaxy, As for the foreground originated from Galactic DWDs, one can study the spatial structure by its anisotropy~\cite{Breivik:2019oar}. In addition, some physical pictures of the early Universe may be hidden in the cosmological \acp{SGWB}~\cite{2000gr.qc.....8027M,Ungarelli:2000jp,2013GWN.....6....4A}.

Specifically, the collection of Galaxy \acp{DWD} could exceed the noise level of space-borne GW detectors~\cite{1990ApJ...360...75H}. 
Instead of forming a background, such a signal is generally classified as a {\it foreground}~\cite{AmaroSeoane:2012je,Nissanke:2012eh,2013ASPC..467...27N,Benacquista:2016dwl}. 
The capability to detect the other isotropic \ac{SGWB} would not be significantly weakened when the foreground is suitably modeled~\cite{Adams:2013qma}. 
Since the strength of foreground is comparable with the detector noise, in the process of data analysis for other signals, the all-sky integrated foreground could be treated as a part of the noise~\cite{Bender:1997hs,Nelemans:2001hp,Barack:2004wc,Edlund:2005ye,Ruiter:2007xx,Nelemans:2009hy}.
%==========TianQin================%

There exist a few programs for the space-borne GW detectors~\cite{Audley:2017drz,Luo:2015ght,Kawamura:2011zz,Crowder:2005nr,Hu:2017mde}. 
In this work, we constrain our focus to TianQin~\cite{Luo:2015ght,Tan:2020xbm,Mei:2020lrl} (and LISA~\cite{Audley:2017drz} when a network is considered). 
TianQin is expected to explore GW astronomy as well as fundamental physics during its operation \cite{Huang:2020rjf,Wang:2019,Liu:2020eko,Fan:2020zhy,Zi:2021pdp,Shi:2019hqa}.
For such detectors, laser interferometry is used to combine three satellites into a six-link triangle GW detector, with which one can build three independent data channels. 
A link is formed when the laser emitted from one satellite is received by the other, which is further used to construct an interferometer.

%==========The methods to detecting SGWB are cross correlation and null channel.================ 

When extracting the GW signals from the resolved sources such as \acp{SBBH}, a common practice is to use the matched filter method~\cite{10.5555/1097023,Owen:1998dk}, where a template bank with various waveforms is used to compare with the data.
However, due to the stochastic nature, the \ac{SGWB} has no definite waveforms. 
Thus, in the realm of \ac{SGWB} detection, the optimal filter method is used instead~\cite{Allen:1997ad}.

For the space-borne GW missions, the laser phase noise is usually orders of magnitude higher than other noises, while the \ac{TDI} can be used to cancel the laser phase noise~\cite{Tinto:2001ii,Tinto:2002de,Hogan:2001jn,Prince:2002hp,Tinto:2004wu}.
The outcome, however, is that instead of the Michelson channel, analysis should be performed on the \ac{TDI} channels, among which the A and E channels are conventional channels, while the T channel is signal insensitive, and is also referred to as the \emph{null channel} or \emph{noise monitoring channel}.

In the search for \acp{SGWB}, one could cross-correlate the outputs from multiple channels~\cite{Hellings:1983fr,Christensen:1992wi,Flanagan:1993ix}.
At first glance, such a method could in principle be applied to the A and E channels.
However, the GW signals projected to the A and E channels belong to orthogonal polarizations.
Therefore, the cross-correlation method cannot be applied on the \ac{TDI} channels~\cite{Adams:2010vc}.
Fortunately, the T channel can provide information about the noise spectrum, and the \ac{SGWB} can be identified in the A and E channels by comparing with the T channel through the \emph{null channel} method~\cite{Robinson:2008fb,Romano:2016dpx}.
%==========In the future, we are able to detect SGWB by TianQin. Combined with LISA,  the cost will also be reduced.================
We calculate the responses of the \ac{SGWB} under different configurations with both the cross correlation and null channel methods, as well as the spectra from various types of sources.

%==========The framework of this article================

The outline of the paper is as follows.
In Sec.~\ref{sec:fundamental} we review the fundamentals of the \ac{SGWB} to provide a theoretical framework for the following calculation.
In Sec.~\ref{sec:methodology} we introduce the methodology to realize the detection of the \ac{SGWB} and show the \ac{PI} sensitivity curves for the different configurations.
In Sec.~\ref{sec:spectrum} we analyze the \acp{SGWB} from different origins.
We summarize the conclusions in Sec.~\ref{sec:summary}.

\section{Basic definitions}\label{sec:fundamental}
\subsection{Fundamentals of the \ac{SGWB}}

The strength of the \ac{SGWB} is described by the ratio of the \ac{SGWB} energy density per logarithmic frequency bin to the critical density of the Universe, $\rho_{\rm c}\equiv3H_{0}^{2}c^{2}/(8\pi G)$~\cite{Allen:1996gp,Schutz:1999xj}:
\be
\label{eq:omega_gw}
\Omega_{\rm gw}(f)=\frac{1}{\rho_{\rm c}}\frac{{\rm d}\rho_{\rm gw}}{{\rm d}(\ln{f})}.
\ee
Here ${\rm d}\rho_{\rm gw}$ is the \ac{SGWB} energy density in the frequency band [$f$, $f+{\rm d}f$],  $H_{0}$ is the Hubble constant, $c$ is the speed of light and $G$ is the gravitational constant.
In this work, we adopt $H_{0}=67.4\,\,\rm{km\,s^{-1}\,Mpc^{-1}}$~\cite{Aghanim:2018eyx}. 
The overall fractional density $\Omega_{\rm gw}$ of the \ac{SGWB} is given by
\be
\label{eq:omega_gw_total}
\Omega_{\rm gw}=\int_{0}^{\infty}{{\rm d}(\ln f)}\,\Omega_{\rm gw}(f).
\ee

The energy density is related to the metric perturbation $h_{ab}(t,\vec{x})$~\cite{Allen:1996vm}:
\be
\label{eq:rho_gw}
\rho_{\rm gw}=\frac{c^{2}}{32\pi G}
\langle \dot{h}_{ab}(t,\vec{x})\dot{h}^{ab}(t,\vec{x})\rangle,
\ee
where $\langle...\rangle$ denotes the ensemble average~\cite{Cutler:1994ys}.
In the plan- wave expansion, the metric perturbation can be expressed as~\cite{Cornish:2001qi}
\be
\label{eq:h_ab}
h_{ab}(t,\vec{x})
=\int_{-\infty}^{\infty}\,{\rm{d}}f\int_{S^{2}}\,{\rm{d}}\Omega_{\hat{k}}
\widetilde{h}_{ab}(f,\hat{k})
e^{i2\pi f(t-\hat{k}\cdot\vec{x}/c)},
\ee
where $\widetilde{h}_{ab}(f,\hat{k})=\sum_{P=+,\times}\widetilde{h}_{P}(f,\hat{k})e_{ab}^{P}(\hat{k})$, with $e^{P}_{ab}(\hat{k})$ being the polarization tensor\cite{Allen:1996gp,Misner:1974qy,2011gwpa.book.....C}.

The detector response can be expressed as the convolution of metric perturbations $h_{ab}(t,\vec{x})$  and the impulse response of the detector $\mathbb{F}^{ab}(t,\vec{x})$~\cite{Romano:2016dpx}, $h(t)=\mathbb{F}^{ab}(t,\vec{x})* h_{ab}(t,\vec{x})$, through which the output in the frequency domain is derived:
\be
\label{eq:hf}
\widetilde{h}(f)
=\int_{S^{2}}\,{\rm{d}}\Omega_{\hat{k}}
\sum_{P=+,\times}\,F^{P}(f,\hat{k})\widetilde{h}_{P}(f,\hat{k})e^{-i2\pi f\hat{k}\cdot\vec{x}/c},
\ee
where $F^{P}(f,\hat{k})=e^{P}_{ab}(\hat{k})F^{ab}(f,\hat{k})$ and
\bea
F^{ab}(f,\hat{k})&=&
\int_{-\infty}^{\infty}{\rm d}\,\tau\int{\rm d}^{3}y\,\mathbb{F}^{ab}(\tau,\vec{y})
e^{-i2\pi f(\tau-\hat{k}\cdot\vec{y}/c)}.
\eea 

Taking the Michelson channel as an example, we have
\be
\label{eq:Dab}
F_{\rm M}^{ab} (f,\hat{k})=\frac{1}{2}
(u_{1}^{a}u_{1}^{b}\mathcal{T}(f,\hat{u}_{1},\hat{k})-
u_{2}^{a}u_{2}^{b}\mathcal{T}(f,\hat{u}_{2},\hat{k})),
\ee
where $\hat{u}_{i}$ are the unit vectors for the two arms of the Michelson channel and $u_i^a$ denotes the $a$ component of the unit vector $\hat{u}_i$. 
$\mathcal{T}(f,\hat{u},\hat{k})$ is the timing transfer function for each arm~\cite{Cornish:2001bb,Cornish:2002rt}. 

The expectation of the cross correlation between the channels $I$ and $J$ is defined as~\cite{Abbott:2007wd}
\be
\label{eq:hstrain}
\langle\widetilde{h}_{I}(f)\widetilde{h}_{J}^{*}(f')\rangle
=\frac{1}{2}\delta(f-f')\Gamma_{IJ}(f)S_{\rm h}(f),
\ee
where $S_{\rm h}(f)$ is the one-sided \ac{PSD} for the \ac{SGWB}, and the \ac{ORF}~\cite{Finn:2008vh} for $I$ and $J$ is
\be
\label{eq:orf_ij}
\Gamma_{IJ}(f)=\frac{1}{8\pi}\sum_{P=+,\times}\int_{S^{2}}{\rm d}\Omega_{\hat{k}}
F^{P}_{I}(f,\hat{k})F^{P*}_{J}(f,\hat{k})
e^{-i2\pi f\hat{k}\cdot\Delta \vec{x}/c},
\ee
where $\Delta \vec{x}=\vec{x}_{I}-\vec{x}_{J}$, and we define $\vec{x}_{I,J}$ as the location where the interference happens. 
The \ac{ORF} reflects the correlation of the response between channels $I$ and $J$, considering a plane wave of frequency $f$ and unit wave vector $\hat{k}$. 
When applied to the same channel, {\it i.e.}, $I=J$, the \ac{ORF} reduces back to the transfer function.

Combining \eqs{eq:omega_gw}--(\ref{eq:hf}) and \eq{eq:hstrain}, the relation between $\Omega_{\rm gw}(f)$ and $S_{\rm h}(f)$ can be derived as~\cite{Nishizawa:2013eqa}
\be
\label{eq:omega2sh}
\Omega_{\rm gw}(f)=\frac{2\pi^{2}}{3H_{0}^{2}}f^{3}S_{\rm h}(f).
\ee

\subsection{Detectors and detector networks}
\begin{figure}[t]
	\centering
	\includegraphics[height=7.5cm]{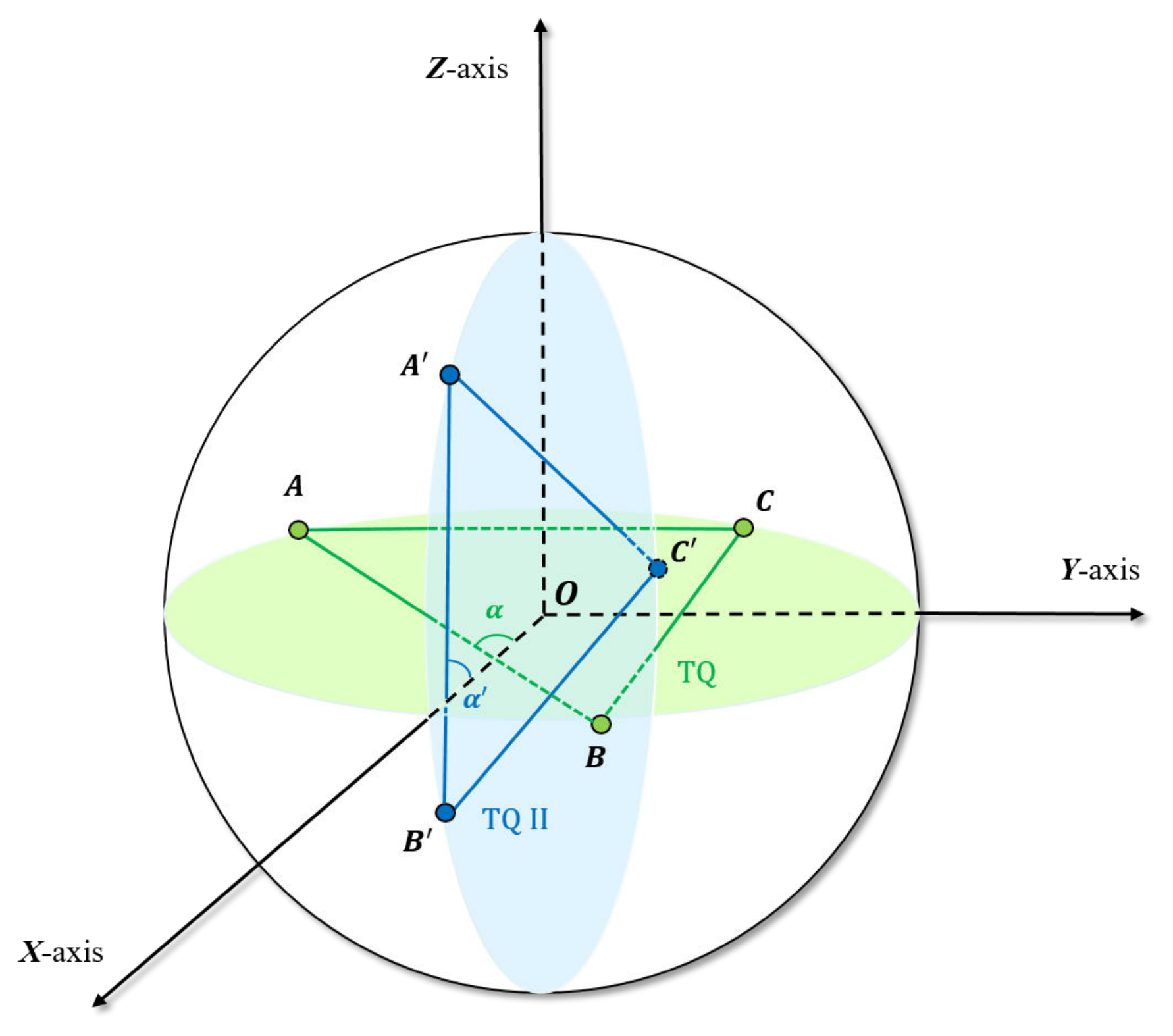}
        \caption{Illustration of TianQin (green) and TianQin II (blue). $\alpha$ and $\alpha'=\alpha+\beta$ are the initial angles of the two constellations.}
	\label{fig:2TQ}
\end{figure}

The nominal working mode for TianQin is set to ``three months on + three months of'': after every three months of the observation, the detector will be down for the next three months~\cite{Ye:2020tze}. 
In \fig{fig:2TQ}, we illustrate the detector coordinate system, where the Earth is placed at the origin. The three satellites A, B, and C (at the vertices) of TianQin run in the {\it X-Y} plane, and we choose the arm length $L_{\rm TQ}=\sqrt{3}\times10^{5}\,\,{\rm km}$, the displacement measurement noise $S_{x}^{1/2}=1\times10^{-12}\,\,{\rm m/Hz^{-1/2}}$ and the residual acceleration noise $S_{a}^{1/2}=1\times10^{-15}\,\,{\rm m\,s^{-2}/Hz^{-1/2}}$, all of which are also listed in \tab{tab:dete_para}. 
Conventionally, the noise level of a GW detector can be represented by its one-sided \ac{PSD} $P_{{\rm n}}(f)$ or the \ac{ASD} $\sqrt{P_{{\rm n}}(f)}$~\cite{Ajith:2006qk}.
The \ac{ASD} of TianQin~\cite{Luo:2015ght,Robson:2019} is shown in \fig{fig:ASD}.
For the Michelson channel, the \ac{ASD} approaches a constant value beyond $f \simeq 10^{-2}~{\rm Hz}$, while the A and E channels oscillate in a sinusoidal way.
The Michelson and A/E channel \acp{PSD} for TianQin can be expressed respectively, as
\bea\label{eq:Pn_Mich}
\nn
P_{\rm n_{M}}(f)&=&\frac{1}{L_{\rm TQ}^{2}}\bigg[S_{x}(f)
+2\bigg(\cos f_{\rm c}^{\rm TQ}
+1\bigg)\\
&\times&\frac{S_{a}(f)}{(2\pi f)^4}\bigg(1+\frac{10^{-4}{\rm Hz}}{f}\bigg)\bigg],
\eea

\bea\label{eq:Pn_AE}
\nn
P_{\rm n_{A,E}}(f)
&=&\frac{1}{L_{\rm TQ}^{2}}\frac{4}{3}\sin^{2} f_{\rm c}^{\rm TQ}
\bigg[\bigg(\cos  f_{\rm c}^{\rm TQ}+2\bigg)S_{x}(f)\\
\nn
+2\bigg(\cos (2f_{\rm c}^{\rm TQ})&+&2\cos f_{\rm c}^{\rm TQ}
+3\bigg)\frac{S_{a}(f)}{(2\pi f)^4}\bigg(1+\frac{10^{-4}{\rm Hz}}{f}\bigg)\bigg],\\
\eea 
where $f_{\rm c}^{\rm TQ}=(2\pi fL_{\rm TQ})/c$.

\begin{figure}[t]
	\centering
	\includegraphics[height=5.5cm]{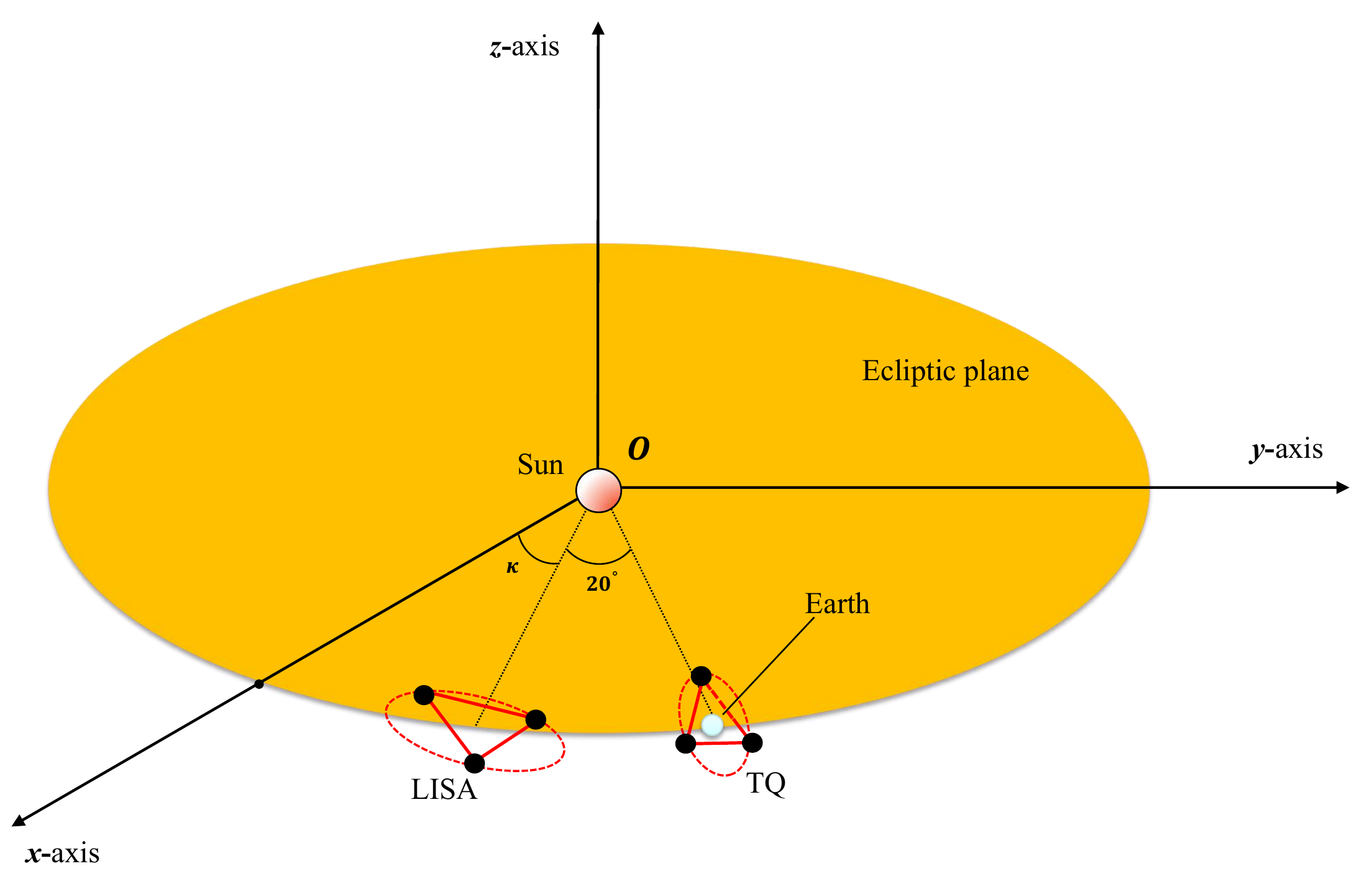}
	\caption{Illustration of configurations of TianQin and LISA. 
		TianQin orbits around the Earth, while LISA orbits the sun trailing the Earth. Notice that the underlying coordinate system differs from that of \fig{fig:2TQ}}
	\label{fig:TQ+LISA}
\end{figure}

\begin{figure}[t]
	\centering
	\includegraphics[height=6cm]{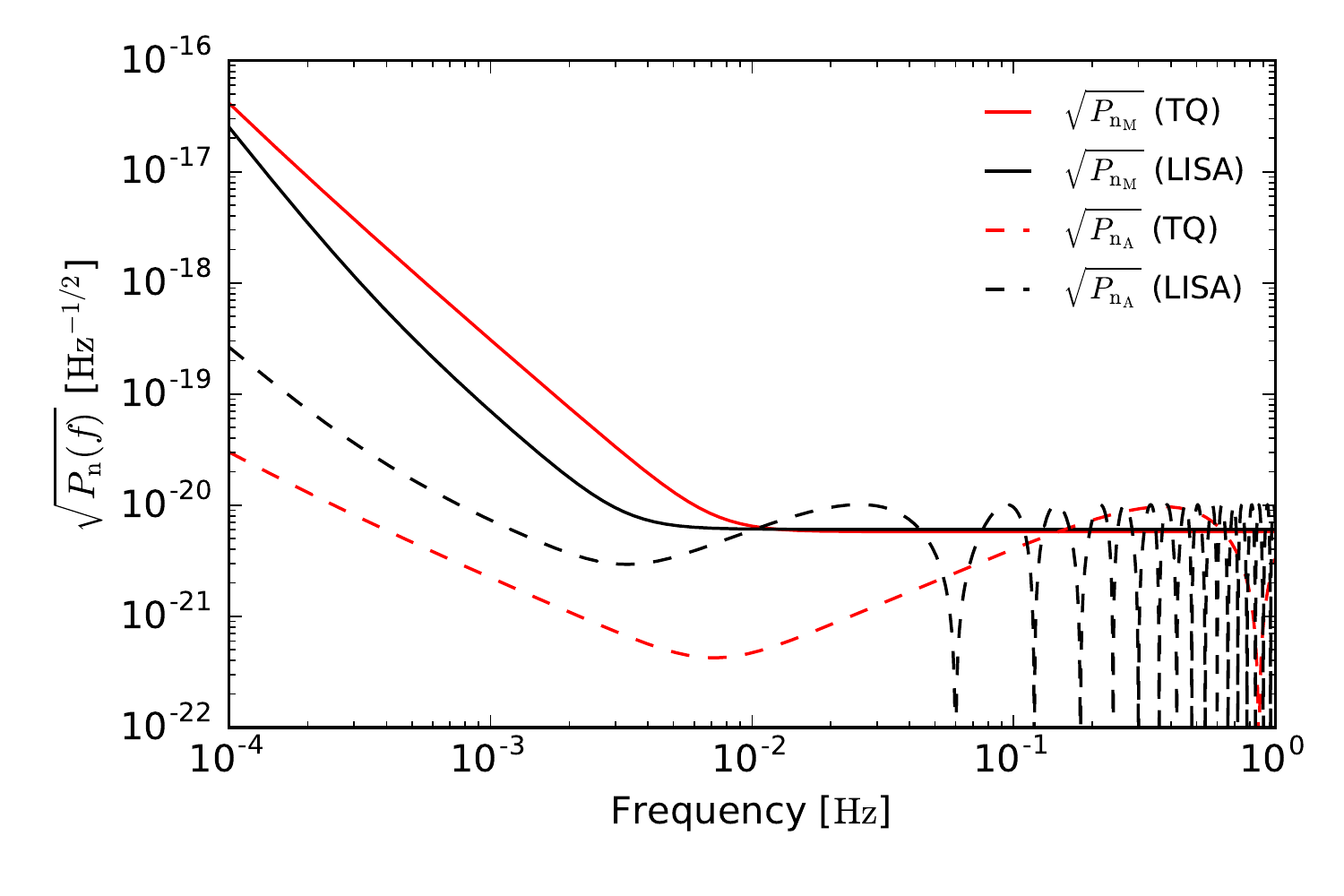}
	\caption{The \ac{ASD} $\sqrt{P_{\rm n}(f)}$ for different channels.The solid and dashed lines denote the Michelson and A(E) channels respectively, where red/black label TianQin/LISA.}
	\label{fig:ASD}
\end{figure}

In addition to individual detectors, we also consider a network of detectors. We start our discussion with TianQin I+II. The nominal working schedule for TianQin II is to prevent long gap between data~\cite{Huang:2020rjf,Wang:2019}. 
As shown in \fig{fig:2TQ}, TianQin II is also composed of three satellites ($\rm{A'}$, $\rm{B'}$, and $\rm{C'}$) orbiting the Earth, an their orbital plane is perpendicular to TianQin's. 
In addition, we also consider the network of TianQin + LISA. 
In \fig{fig:TQ+LISA}, we illustrate the orbits of TianQin and LISA in the ecliptic coordinate system. 
Solar System barycenter is located at the center of the coordinate system and the ecliptic plane is chosen as the {\it x-y} plane in which both TianQin and LISA revolve around the Sun at the same rate that the Earth moves, keeping a fixed separation of $\sim 20^{\circ}$ in the orbit of the Earth. 
The corresponding parameters of LISA are also listed in \tab{tab:dete_para}, and the \ac{ASD} of LISA is shown in \fig{fig:ASD}. As a shorthand notation, in figures and equations, we may use TQ for TianQin, TQ II for TianQin II, TT for TianQin I+II, TL for TianQin + LISA, and TTL for TianQin I+II + LISA.

\begin{table}
	\begin{center}
		\caption{Basic parameters of TianQin\cite{Luo:2015ght} and LISA\cite{Robson:2019}, where we list the arm length $L$, the displacement measurement noise $S_{x}^{1/2}$ and the residual acceleration noise $S_{a}^{1/2}$.}\label{tab:dete_para}
		\begin{tabular}{*{3}{|c}|}
			\hline
			Parameter            & TianQin             & LISA\\
			\hline
			$L$          & $\sqrt{3}\times10^{5}\,\,{\rm km}$    & $2.5\times10^{6}\,\,{\rm km}$ \\
			\hline
			$S_{x}^{1/2}$ & $1\times10^{-12}\,\,{\rm m/Hz^{1/2}}$   & $1.5\times10^{-11}\,\,{\rm m/Hz^{1/2}}$ \\
			\hline 
			$S_{a}^{1/2}$ & $1\times10^{-15}\,\,{\rm m\,s^{-2}/Hz^{1/2}}$ & $3\times10^{-15}\,\,{\rm m\,s^{-2}/Hz^{1/2}}$\\
			\hline
		\end{tabular}
	\end{center}
\end{table}

\section{Methodology}\label{sec:methodology}

The output $s(t)$ of each channel in a detector generally consists of the channel noise $n(t)$ and the signal response $h(t)$ in the channel from the \ac{SGWB}~\cite{Finn:1992wt}. 
The key point in \ac{SGWB} detection is to distinguish the signal from the noise.
The stochastic nature of SGWB means that one can not apply the mathced filtering method.
Meanwhile, it is very weak, so that it is nearly impossible to extract the SGWB with only one channel
~\cite{Bose:2005fm,Whelan:2005sk,Abbott:2009ws,Regimbau:2014uia,Aasi:2014zwg,Aasi:2014jkh,Meacher:2015iua,Nishizawa:2016jst,TheLIGOScientific:2016dpb,LIGOScientific:2019vic,Seto:2020mfd,Abbott:2021xxi}. 
However, once feasible strategies are adopted, the \ac{SGWB} can be detectable even when the signal is weaker than the noise. 
Such strategies include performing a correlation between two (or more) independent channels (called the ``cross correlation''~\cite{10.1093/mnras/227.4.933,Christensen:1992wi,Flanagan:1993ix,Vitale:1996xv,Allen:1996sw,Rakhmanov:2008is}) and constructing the null channel~\cite{Robinson:2008fb,Romano:2016dpx} in a single detector. 
We shall discuss these two methods in detail and present the formula of the \ac{SNR} in parallel.

%%%%%%%%%%%%%%%%%
\subsection{Cross-correlation method}
The cross-correlation method for \ac{SGWB} detection requires at least two noise-independent channels~\cite{Dhurandhar:2002zcl}.
Although a single triangle-shaped detector contains multiple noise-independent channels, the two channels are also signal-independent~\cite{Cutler:1997ta,Adams:2010vc}.
Therefore, one cannot use one triangle-shaped GW detector to detect the \ac{SGWB} using the cross-correlation method.

To perform a cross correlation, one needs to adopt two channels $s_I$ and $s_J$, in which the only related items are the signals~\cite{Meacher:2014aca}:

\bea
\nn
\label{s_cross}
s_{I}(t)&=&n_{I}(t)+h_{I}(t)\\
s_{J}(t)&=&n_{J}(t)+h_{J}(t).
\eea
Here $n(t)$ represents the channel noise and $h(t)$ stands for the GW signal.
The noise-independence condition implies that $\langle n_I(t)n_J(t)\rangle=0$.
If the noises are stationary, we have~\cite{Sathyaprakash:1991mt}
\bea
\nn
\langle \widetilde{n}_{I}(f)\widetilde{n}_{I}^{*}(f')\rangle&=&\frac{1}{2}\delta(f-f')P_{{\rm n}_{I}}(f)\\
\langle \widetilde{n}_{J}(f)\widetilde{n}_{J}^{*}(f')\rangle&=&\frac{1}{2}\delta(f-f')P_{{\rm n}_{J}}(f),
\eea
where $P_{{\rm n}_{I,J}}(f)$ is the \ac{PSD} for the noise of channels $I$ and $J$.

Since the noise and the \ac{SGWB} signal are statistically independent of each other, the expectation value of the cross correlation for two channel outputs is entirely contributed by the \ac{SGWB} signal:
%\be
%\langle s_I (t) s_J (t) \rangle = \langle h_I (t) h_J (t) \rangle
%\ee
\bea
\label{eq:Scor}
\nn
\langle S_{IJ} \rangle &\equiv& \int^{T/2}_{-T/2}{\rm d}t 
\int^{\infty}_{-\infty}{\rm d}f\int^{\infty}_{-\infty}{\rm d}f'
\langle \widetilde{h}_I (f) \widetilde{h}^{*}_J (f')\rangle  \\
&\times&\frac{\Gamma_{IJ}(|f|)S_{\rm h}(|f|)}{P_{{\rm n}_{I}}(|f|)P_{{\rm n}_{J}}(|f|)}e^{-i2\pi (f-f')t},
\eea
where $S_{\rm h}(f)$ is defined in \eq{eq:hstrain} and related to the energy spectrum density of the \ac{SGWB} via \eq{eq:omega2sh}.

Under the assumption that the intrinsic noises of the detectors are significantly larger in magnitude than the \ac{SGWB} signal, $|n_{I,J}(t)| \gg |h_{I,J} (t)|$, the noise component can be estimated by the variance of the correlation 
$\sigma^2 \equiv \langle S^2_{IJ} \rangle- \langle S_{IJ} \rangle^2$.
As a result, the expected \ac{SNR} of the correlated \ac{SGWB} signals from cross correlating two detectors is given by~\cite{Sharma:2020btq}
\bea
\label{eq:snr_cross}
{\rm SNR} \equiv {\langle S_{IJ} \rangle \over \sigma}
&=&\sqrt{2\,T_{\rm tot}\int_{f_{\rm min}}^{f_{\rm max}} {\rm d}f\,
\frac{|\Gamma_{IJ}(f)|^{2}S^{2}_{\rm h}(f)}{P_{{\rm n}_{I}}(f)P_{{\rm n}_{J}}(f)}},
\eea
where $f_{\rm min}$ and $f_{\rm max}$ denote the minimum and maximum observation frequencies of the channels. The prefactor $T_{\rm tot}$ is the total time that two detectors coincidently operate.
Therefore, the longer two detectors operate jointly, the larger \ac{SNR} it will accumulate.

\subsubsection{{\rm TianQin I+II}}

We first consider TianQin I+II when adopting the cross correlation method.
The nominal working mode implies that the two TianQin observatories have no common operation time, which makes the cross correlation inapplicable. 
Therefore, the aim of detecting the \ac{SGWB} with TianQin I+II requires an extension of the current nominal operation time which brings extra constraints to the design of the satellites.

Without loss of generality, suppose that at time $t$ the unit vectors for the three arms of TianQin are as constructed in \fig{fig:2TQ},
\bea
\nn
{\rm AB}: \,\, \hat{u}_{1}(\alpha)&=&\big(\cos\alpha(t),\sin\alpha(t),0\big)\\
\nn
{\rm AC}: \,\,\hat{u}_{2}(\alpha)&=&\big(\cos(\alpha(t)+\pi/3),\sin(\alpha(t)+\pi/3),0\big)\\
\nn
{\rm BC}: \,\,\hat{u}_{3}(\alpha)&=&\big(\cos(\alpha(t)+2\pi/3),\sin(\alpha(t)+2\pi/3),0\big),\\
\eea
and those for TianQin II are
\bea
\nn
{\rm A'B'}: \,\,\hat{u}'_{1}(\alpha')&=&\big(\cos\alpha'(t), 0, \sin\alpha'(t) \big)\\
\nn
{\rm A'C'}: \,\, \hat{u}'_{2}(\alpha')&=&\big(\cos(\alpha'(t)+\pi/3),0,\sin(\alpha'(t)+\pi/3)\big)\\
\nn
{\rm B'C'}: \,\, \hat{u}'_{3}(\alpha')&=&\big(\cos(\alpha'(t)+2\pi/3),0,\sin(\alpha'(t)+2\pi/3)\big).\\
\eea

Since the noises of TianQin and TianQin II are independent, there are multiple ways to construct two channels for cross correlation.
As a start, we focus on the study of two Michelson channels.
Each Michelson channel can be constructed by connecting two adjoint links (following the standard construction) or by a combination of six links. 
In \fig{fig:2TQ}, a Michelson channel can be built by connecting the links $\rm{AB}$ and $\rm{AC}$ of TianQin as well as the other produced by the links $\rm{A'B'}$ and $\rm{A'C'}$ of TianQin II. 
The response functions in these two Michelson channels are
\bea
\nn
{\rm TQ}: &\quad F_{\rm M_{1}}^{P}(f,\hat{k},\alpha)&=e^{P}_{ab}(\hat{k})F_{\rm M}^{ab} (\hat{u}_1,\hat{u}_2,\hat{k},f) \\
{\rm TQ\,II}: &\quad F_{\rm M_{2}}^{P}(f,\hat{k},\alpha,\gamma_0)&=e^{P}_{ab}(\hat{k})F_{\rm M}^{ab} (\hat{u}'_1,\hat{u}'_2,\hat{k},f),  \label{eq:F1F2}
\eea
where the explicit form of $F_{\rm M}^{ab}$ is defined in \eq{eq:Dab}. Substituting \eq{eq:F1F2} into \eq{eq:orf_ij}, the \ac{ORF} for two Michelson channels from TianQin and TianQin II, respectively, can be written as
\bea
\label{eq11}
\nn
\Gamma^{'\rm TT}_{\rm M_{1}M_{2}}(f,\alpha,\gamma_0)
&=&\frac{1}{8\pi}\sum_{P=+,\times}\int_{S^{2}}{\rm d}
\Omega_{\hat{k}}F^{P}_{\rm M_1}(f,\hat{k},\alpha)\\
&\times&F^{P*}_{\rm M_2}(f,\hat{k},\alpha,\gamma_0)
e^{-i2\pi f\hat{k}\cdot \Delta \vec{x}_{\rm TT}/c},
\eea
where $\Delta \vec{x}_{\rm TT}=\overrightarrow{A'A}$, and $\gamma_{0}(t)\equiv \alpha'(t)-\alpha(t)$.

The instantaneous \ac{ORF} is dependent on the initial angles $\alpha$ and $\gamma_0$.
We can define a time-averaged \ac{ORF} that is free of the angle dependence,
\be
\label{eq:gamma_m1m2}
\widehat{\Gamma}^{\rm TT}_{\rm M_{1}M_{2}}(f,\gamma_0)
=\frac{1}{2\pi}\sqrt{\int_{0}^{2\pi}{\rm d}\alpha\,
\big|\Gamma^{'\rm TT}_{\rm M_{1}M_{2}}(f,\alpha,\gamma_0)\big|^{2}}.
\ee
Notice that for either TianQin I or II, two orthogonal channels can be constructed, which we call as $\rm M_{1}$, $\rm M_{2}$, $\rm M'_{1}$, and $\rm M'_{2}$, and one can thus build the total \ac{ORF} of TianQin I+II as \cite{Seto:2020mfd}
\bea
\label{eq:gamma_total}
\nn
\widehat{\Gamma}^{\rm TT}_{\rm MM}(f,\gamma_0)
&=&\bigg[\big|\widehat{\Gamma}^{\rm TT}_{\rm M_{1}M_{2}}(f,\gamma_0)\big|^{2}
+\big|\widehat{\Gamma}^{\rm TT}_{\rm M'_{1}M_{2}}(f,\gamma_0)\big|^{2}\\
&+&\big|\widehat{\Gamma}^{\rm TT}_{\rm M_{1}M'_{2}}(f,\gamma_0)\big|^{2}
+\big|\widehat{\Gamma}^{\rm TT}_{\rm M'_{1}M'_{2}}(f,\gamma_0)\big|^{2}\bigg]^{1/2}
\eea

\begin{figure}[t]
	\centering
	\includegraphics[height=5.5cm]{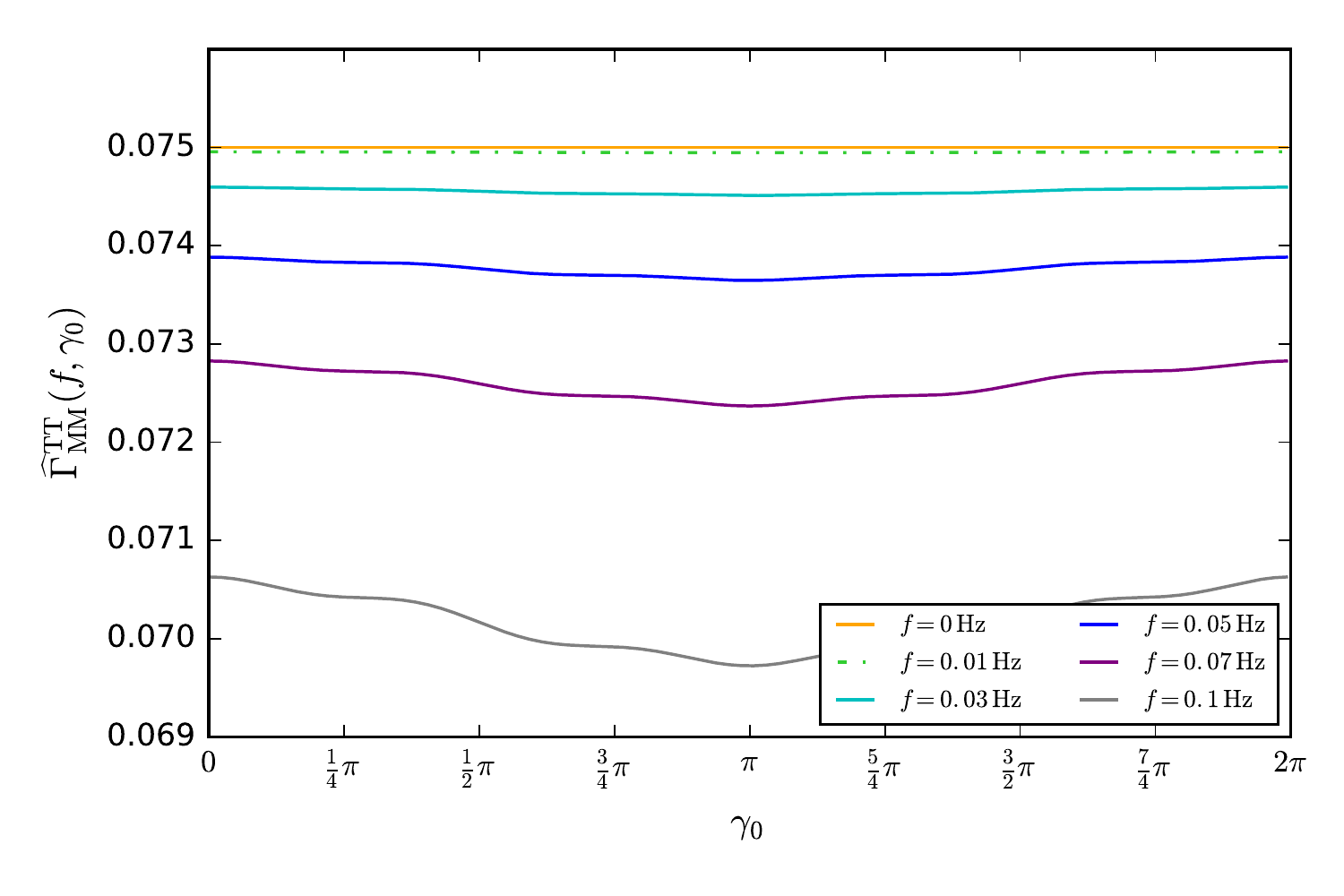}
        \caption{Dependence of the total \ac{ORF} $\widehat{\Gamma}^{\rm TT}_{\rm MM}(f,\gamma_0)$ (\eq{eq:gamma_total}) on $\gamma_0$. The optimal \ac{ORF} of TianQin I+II can be obtained after adopting $\gamma=2\,n\pi$. We ignore the \ac{ORF} with frequencies beyond $0.1\,\,\rm Hz$.}
	\label{fig:ORF_beta}
\end{figure}

\begin{figure}[t]
	\centering
	\includegraphics[height=6cm]{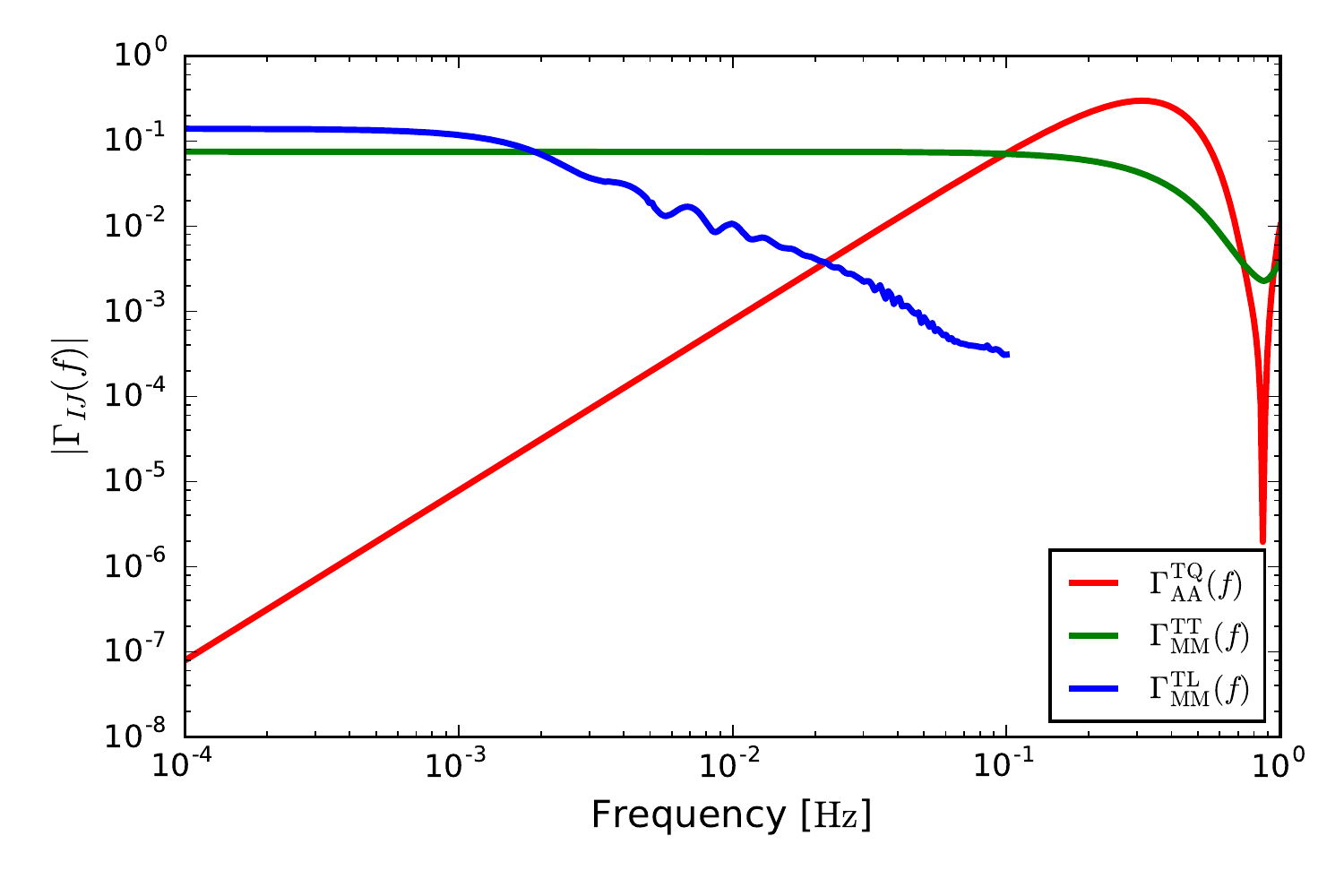}
        \caption{{Response function or \ac{ORF} for different constructions. (i) The red line assumes the null channel for \ac{TDI} channels (AE) of TianQin.
        (ii) The green line is the \ac{ORF} of TianQin I and II, for which we use the Michelson channel and cross-correlation method.
        (iii) The blue line represents the situation when we combine the Michelson channels of TianQin and LISA by adopting cross correlation. 
         We truncate the blue line at 0.1 Hz as due to the limitation of numerical calculation accuracy.}}
	\label{fig:ORF_3cases}
\end{figure}

As shown in \fig{fig:ORF_beta}, in the low-frequency range the total \ac{ORF} is independent of $\gamma_0$. However, it exhibits a periodic pattern at higher frequencies. The total \ac{ORF} reaches the optimal performance at $\gamma_0=2\,n\pi, n=0,1,2,3,\dots$:
\be
\Gamma^{\rm TT}_{\rm MM}(f)=\widehat{\Gamma}^{\rm TT}_{\rm MM}(f,\gamma_0)|_{\gamma_0=2n\pi}.
\ee
This suggests that if we aim to optimize the SGWB detection capability, 
TianQin and TianQin II should be launched in a way that their initial phases differ by the degree $\gamma_0=\alpha-\alpha=2\,n\pi$ (shown in \fig{fig:2TQ}), which sets the constraints for the design of TianQin I+II: the interval that TianQin II enters the orbit should be $n\tau$ after that of TianQin, with $\tau$ being the orbital period for TianQin. 
We assume the optimal \ac{ORF} in the following analysis.
The optimal \ac{ORF} for the present configuration, $|\Gamma^{\rm TT}_{\rm MM}(f)|$ is shown by the green line in \fig{fig:ORF_3cases}, where the optimal \ac{ORF} is 3/40 when $f\ll f_*^{\rm TT}\simeq0.28\,\,\rm Hz$.
 
\subsubsection{{\rm TianQin + LISA}}
We now extend the study to a network of  TianQin + LISA, whose configuration is also shown in  \fig{fig:TQ+LISA}. 
According to the time lines of two projects, their are planed to operate almost at same time, with LISA working continuously. 
The position vector of the $n$th ($n=1,2,3$) TianQin satellite can be decomposed into two parts:
\be
\vec{x}_n(t)=\vec{X}(t)+\vec{\tilde{x}}_n(t),
\ee
where $\vec{X}(t)$ represents the position of the Earth and $\vec{\tilde{x}}_n(t)$ accounts for the motion around the Earth. The components of $\vec{x}_n(t)$ and $\vec{x}'_n(t)$ in a common coordinate system for TianQin and the LISA are given in the Appendix, and their full expressions can be found in~Refs.~\cite{Hu:2018yqb} and~\cite{Cornish:2002rt}, respectively.

The unit vectors for the three arms of TianQin and LISA are
\bea
\nn
\hat{u}_n(t)&=&\epsilon_{nml} (\vec{x}_m(t)-\vec{x}_l(t))/L_{\rm TQ}\\
\hat{v}_n(t)&=&\epsilon_{nml}  (\vec{x}'_m(t)-\vec{x}'_l(t))/L_{\rm LISA},
\eea
and the response functions for the two Michelson channels in TianQin and LISA are, respectively,
\bea
\nn
{\rm TQ}:&\quad F_{\rm M_1}^{P}(f,\hat{k},t)&=e^{P}_{ab}(\hat{k})F_{\rm M}^{ab} (\hat{u}_1(t),\hat{u}_2(t),\hat{k},f) \\
\nn
{\rm LISA}:&\quad F_{\rm M_2}^{P}(f,\hat{k},t)&=e^{P}_{ab}(\hat{k})F_{\rm M}^{ab} (\hat{v}_1(t),\hat{v}_2(t),\hat{k},f).  \label{eq:F11F22}\\
\eea
The \ac{ORF} of the above two channels is 
\bea
\nn
\Gamma^{\rm 'TL}_{\rm MM} (f,t)
&=&\frac{1}{8\pi}\sum_{P=+,\times}\int_{S^{2}}{\rm d}\Omega_{\hat{k}}F^{P}_{\rm M_1}(f,\hat{k},t)\\
&\times&F^{P*}_{\rm M_2}(f,\hat{k},t)
e^{-i2\pi f\hat{k}\cdot \Delta \vec{x}_{\rm TL}/c}.
\eea

Although the position vectors for the TianQin satellites [$\vec{x}_n(t)$] and LISA [$\vec{x}'_n(t)$] involve the time-dependent angle [$\alpha_{\rm TQ}(t)$ and $\alpha_{\rm LISA}(t)$], they are periodic, which leads to periodic behavior in $F_{\rm M1}^{P}$ and $F_{\rm M2}^{P}$ as well as $\Delta \vec{x}_{\rm TL}$. Therefore, the \ac{ORF} oscillates periodically and the overall performance within one period can be obtained by taking a time average, 
\be
\Gamma^{\rm TL}_{\rm M_1M_2}(f)=
\frac{1}{\tau}\sqrt{\int_{0}^{\tau}{\rm d}t\,
\left[\Gamma^{\rm 'TL}_{\rm M_1M_2}(f,t)\right]^2},
\ee
where $\tau$ is the orbital period of the Earth. 
Similar to the TianQin I+II configuration [{i.e.,} \eq{eq:gamma_total}], we calculate the total \ac{ORF} $\Gamma^{\rm TL}_{\rm M}(f)$ for TianQin-LISA joint network, which is also shown by the blue line in \fig{fig:ORF_3cases} and truncated at 0.1 Hz.
The averaged \ac{ORF} of TianQin + LISA is almost fixed around the value of $0.14$ as $f\ll f_*^{\rm TL} ={\rm min}[c/(2\pi|\Delta \vec{x}_{\rm TL}|),c/(2\pi L_{\rm TQ}),c/(2\pi L_{\rm LISA})]\simeq 1\,\,\rm mHz$, where we have adopted a LISA arm length of $L_{\rm LISA}=2.5\times10^{6}\,\,\rm km$~\cite{Audley:2017drz}.

%==================The result comparison between Seto's and ours==============%
Before we move on, we highlight the comparison of our results with Ref.~\cite{Seto:2020mfd}, where the \ac{ORF} for TianQin I+II and TianQin + LISA were also calculated and presented. 
There are slight differences in the chose parameterizations. 
1) Ref.~\cite{Seto:2020mfd} chose the normalized \ac{ORF}, which differs from our un-normalized \ac{ORF} by a prefactor of 5.
2) An interferometer with the two arms seperated by $60^{\circ}$ is equivelant to an L-shaped interferometer with arm length shortened by $\sin(60^\circ)$ \cite{Cutler:1997ta}, this factor is included in our presentation, but is not in Ref.~\cite{Seto:2020mfd}.
Under the long-wavelength approximation required by Ref.~\cite{Seto:2020mfd}, we confirm that we can accurately reproduce the results of Ref.~\cite{Seto:2020mfd}, and within numerical error, while our results does not rely on the long-wavelength approximation and thus they are still valid at higher frequencies.

\subsection{Null channel method}
An alternative method is the null channel method, which is applicable to \ac{SGWB} detection by TianQin or LISA alone~\cite{Tinto:2003uk,Tinto:2003vj,Vallisneri:2004bn,Tinto:2004yw,Adams:2010vc}. 
The key point is to construct three orthogonal channels where the \ac{SGWB} signal in one of the channels is highly suppressed. 
The T channel constructed through \ac{TDI} methods can do just that, with the A/E channels being the signal-sensitive channels.
The outputs of the \ac{TDI} channels can be written as~\cite{Robinson:2008fb}
\bea
\nn
s_{\rm T}(t)&\simeq&n_{\rm T}(t),\quad f\lesssim f_{*}\\
s_{\rm A,E}(t)&=&n_{\rm A,E}(t)+h_{\rm A,E}(t).
\eea

To extract the \ac{SGWB} signal from the channel noise, the \ac{TDI} channels should be processed to form a new output~\cite{Smith:2019wny},
\be
s_{0}(t,t')=\sum_{I=\rm A, E}\left[s_{I}(t)s_{I}(t')-\langle n_{I}(t)n_{I}(t')\rangle\right],
\ee
as well as the autocorrelation of the new outputs
\bea
\label{eq:K}
\nn
K&=&\int_{-T/2}^{T/2}{\rm d}t\int_{-T/2}^{T/2}{\rm d}t'
s_{0}(t,t')Q_{II}(t-t')\\
\nn
&\approx&
\int_{-T/2}^{T/2}{\rm d}t
\int_{-\infty}^{\infty}{\rm d}f\int_{-\infty}^{\infty}{\rm d}f'
\,s_{0}(f,f')\\
&\times&
\frac{\Gamma_{II}^{\rm TQ}(|f|)S_{\rm h}(|f|)}{P_{{\rm n}_{I}}^{2}(|f|)}e^{-i2\pi (f'-f)t}.
\eea
We assume that
\be
P_{{\rm n}_{I}}(f)=z_{I}(f)P_{\rm n_{T}}(f),
\ee
where the component $z_{I}(f)$ will have no influence on the result of the \ac{SNR} calculation. The precise definition of $z_{I}(f)$ requires the accurate modeling of the channel noises and the real-time simulation data, which are beyond the scope of this article.

By the construction in \eq{eq:K}, the \ac{SGWB} signal contributes to the ensemble average of  $\langle K\rangle$. On the other hand, the variance is $\sigma^2=\langle K^2 \rangle-\langle K \rangle^2\simeq\langle K^2\rangle$, under the assumption that $h(t)\ll n(t)$. The \ac{SNR} $\rho$ of the null channel for the \ac{SGWB} detection is~\cite{Smith:2019wny}
\be
\label{eq:snr_null0}
\rho
=\frac{\langle K\rangle}{\sigma}\simeq \sqrt{\sum_{I=\rm A,E}T_{\rm tot}\int^{f_{\rm max}}_{f_{\rm min}}{\rm d}f\,
\bigg(\frac{\Gamma^{\rm TQ}_{II}(f)S_{\rm h}(f)}{P_{{\rm n}_{I}}(f)}\bigg)^{2}},
\ee
where $T_{\rm tot}$ is the observation time. It is straightforward to obtain the transfer function $\Gamma^{\rm TQ}_{\rm AA}(f)$ for TianQin through \eq{eq:orf_ij}\footnote{In some studies (e.g., Ref.~\cite{Lu:2019log}), the definition of the transfer function can differ from our definition by a constant numerical factor.}. 
In the ideal symmetric scenario, the \acp{PSD} and the transfer functions of A and E channels take the same form~\cite{Krolak:2004xp}. 
In this case, \eq{eq:snr_null0} reduces to:
\be
\label{eq:snr_null}
\rho
=\sqrt{2\,T_{\rm tot}\int_{f_{\rm min}}^{f_{\rm max}}{\rm d}f\,
\bigg(\frac{\Gamma^{\rm TQ}_{\rm AA}(f)S_{\rm h}(f)}{P_{\rm n_{\rm A}}(f)}\bigg)^{2}},
\ee
The response function $\Gamma^{\rm TQ}_{\rm AA}$ adopted can be found as the red line in \fig{fig:ORF_3cases}.
The A/E channels are constructed as the difference between two Michelson channels delayed in time by $2L/c$; therefore, for a frequency $f\to0$, {i.e.,} a GW wavelength $\lambda\gg L$, the difference vanishes and one has $\Gamma^{\rm TQ}_{\rm AA}(f)\to0$. 

We remark that the general assumption that the background is much weaker than the noise is not universal.
When the condition is invalid, it is no longer appropriate to calculate the \ac{SNR} with \eqs{eq:snr_cross} or (\ref{eq:snr_null}).
The modified expression for the \ac{SNR} is~\cite{Cornish:2001bb,Kudoh:2005as}
\bea
\label{eq:snr_modify}
\nn
{\rm SNR} \equiv {\langle S_{IJ} \rangle \over \sigma}
&=&\sqrt{2T_{\rm tot}\int_{f_{\rm min}}^{f_{\rm max}} {\rm d}f\,
\frac{\Gamma^{2}_{IJ}(f)S^{2}_{\rm h}(f)}
{P_{{\rm n}_{I}}(f)P_{{\rm n}_{J}}(f)W_{\rm C}(f)}},\\
\eea
where the correction term $W_{\rm C}(f)$ is 
\bea
\nn
W_{\rm C}(f)&=&1+\frac{S_{\rm h}(f)\big[\Gamma_{II}(f)P_{{\rm n}_{I}}(f)+\Gamma_{JJ}(f)P_{{\rm n}_{J}}(f)\big]}{P_{{\rm n}_{I}}(f)P_{{\rm n}_{J}}(f)}\\
&+&\frac{\big[\Gamma_{IJ}(f)S_{\rm h}(f)\big]^{2}
\bigg[1+\frac{\Gamma_{II}(f)\Gamma_{JJ}(f)}{\Gamma^{2}_{IJ}(f)}\bigg]}
{P_{{\rm n}_{I}}(f)P_{{\rm n}_{J}}(f)}.
\eea
Notice that this expression is versatile and can be extended to the null channel methodology by setting $I=J$.

From \eqs{eq:snr_cross} and (\ref{eq:snr_null}), one can conclude that no matter what method is adopted, the \ac{SNR} for \ac{SGWB} detection is proportional to $\sqrt{T_{\rm tot}}$. 
However, based on the above results, it is not straightforward to determine which configuration is the best for \ac{SGWB} detection. 
In the next section, we will assess the \ac{SGWB} detection capabilities of different configurations.

\subsection{Power-law integrated sensitivity curve}
There is no straightforward way to visually compare the \ac{SGWB} strength to a detector sensitivity.
A convenient way of illustrating the sensitivity is to assume a power-law spectrum for \ac{SGWB} $\Omega_{\rm gw}(f)=\Omega_{\epsilon}(f/f_{\rm ref})^{\epsilon}$\footnote{Readers are reminded that this form does not imply an Einstein summation}~\cite{Thrane:2014yza}, and present the \emph{power-law integrated sensitivity curve}.
There are only two free parameters in this expression: the reference frequency $f_{\rm{ref}}$ can be arbitrary, and $\Omega_{\epsilon}$ is related to $\epsilon$. 
One can thus define the \ac{PI} sensitivity curve $\Omega_{\rm PI}(f)={\rm max}_{\epsilon}[\Omega_{\epsilon}(f/f_{\rm ref})^{\epsilon}]$, where the power-law energy density is defined as~\cite{Thrane:2013oya}
\be
\Omega_{\epsilon}=\rho_0\left[2T_{\rm tot}\int_{f_{\rm min}}^{f_{\rm max}}{\rm d}f
\frac{(f/f_{\rm ref})^{2\epsilon}}{\Omega_{\rm n}^{2}(f)}\right]^{-1/2},
\ee
and $\epsilon$ is chosen as $-10,\ -9,\ \ldots,\ 9,\ 10$, 
$\rho_0$ is the \ac{SNR} threshold, and the corresponding energy density is defined as

\be
\Omega_{\rm n}(f)=
\frac{2\pi^{2}}{3H_{0}^{2}}f^{3}\frac{\sqrt{P_{{\rm n}_{I}}(f)P_{{\rm n}_{J}}(f)}}{\Gamma_{IJ}(f)},
\ee	
where $I\neq J$ and $I=J$ are related to the cross correlation and the null channel, respectively.

\begin{figure}[t]
	\centering
	\includegraphics[height=6cm]{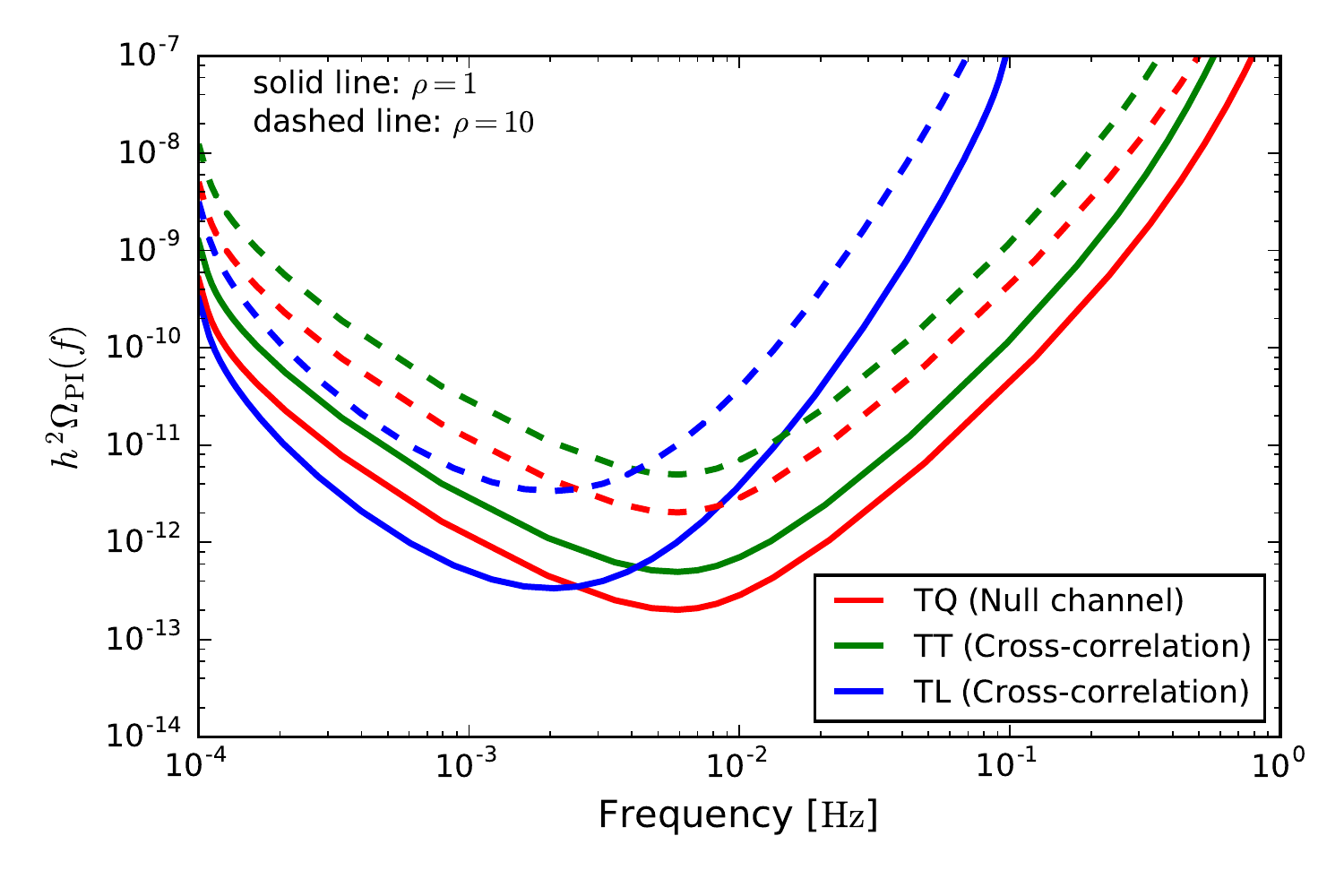}
        \caption{{PI} sensitivity curves for various detector configurations. The operation time is set to $T_{\rm op}=1\,\,{\rm yr}$ and the \ac{SNR} threshold $\rho_{0}=1,10$. We expect 0.5 years of observation time per year for TianQin (and TianQin + LISA network). Since the designed operation scheme of TianQin I + II does not permit joint observation, in this work we assume a modified working schedule to support a joint observation of 4 months per year.}
	\label{fig:PI}
\end{figure}

In \fig{fig:PI} we present the \ac{PI} sensitivity curves under different configurations, assuming an operation time $T_{\rm op}=1\,\,\rm yr$.
Notice that due to the working mode of TianQin, the nominal observation time for the TianQin-LISA joint network $T_{\rm tot}$ equals to half of the operation time $T_{\rm op}$. 
For TianQin and TianQin II, the nominal design does not contain overlap in observation time, and we have to discuss on basis of a different observation scenario, which we choose to consider a ``four months on + two months off'' scheme, therefore, one can expect four months of joint observation per year.
 
%============The factors that affecting the PI sensitivity curve============%
In order to make a fair comparison of the detection capabilities of different configurations, one has to take several factors into account.
From a superficial glance at \fig{PI_curve_3config}, one might jump into the false conclusion that the single TianQin possesses a better sensitivity for the \ac{SGWB} detection, especially in the frequency band beyond a few $\rm mHz$. 
This is only partially true: the most ideal case for detecting an \ac{SGWB} using the cross-correlation method would involve multiple detectors that have maximum correlation in signal and zero correlation in noise, and the maximum correlation in signal is only achieved when the two detectors are colocated and cooriented.
Therefore, for a general multidetector network, a finite difference in orientation and location between detectors reduces their ability to detect the \ac{SGWB}.
Second, a simultaneous operation between multiple detectors is critical for the \ac{SGWB} detection, and the total simualtenous operation time $T_{\rm tot}$ is always shorter than a single detector's operation time.
These two factors reduce the sensitivity for the cross correlation method compared with the null channel method.
However, recall that in order for the null channel method to succeed, one has to rely heavily on the detailed understanding of the detector noises, while the cross-correlation method represents a much more feasible approach. 
Therefore, conceptually, the cross-correlation method is much more robust and reliable than the null channel method.

%==============The additional dicussion for the PI sensitivity curve================%
Finally, we remark that the there are multiple ways to present the detectors' sensitivity to \ac{SGWB}.
In addition to the \ac{PI} curve we adopt in this work, some might adopt a peak-integrated sensitivity curve that assumes a peak-shaped form for the \ac{SGWB} spectrum~\cite{Schmitz:2020syl}. 
Furthermore, by showing in \ac{PI} curve, one gains the advantage of assessing \ac{SNR} straightforwardly.
In this formalism, the \ac{PI} curve drops when observation time increases~\cite{Allen:1997ad,Moore_2014}.
On the other hand, some studies have preferred to present the \ac{SGWB} in a time-independent and threshold-independent style through the corresponding energy density $\Omega_{\rm n}$~\cite{AmaroSeoane:2012je,Caprini:2019egz,Giovannini:2019oii,Poletti:2021ytu}. 

%%%%%%%%%%%%%%%%%%%%
\section{Spectrum of the \ac{SGWB}}\label{sec:spectrum}
%%%%%%%%%%%%%%%%%%%%%

In this section we briefly review a number of the potential sources for the \ac{SGWB} in the $\rm{mHz}$ range. The sources can be divided into two classes, namely, astrophysical \ac{SGWB} and the cosmological \acp{SGWB}, and we study their strengths and spectral shapes. The former contains inspirals of the \acp{DWD}, \acp{BNS}, and \acp{SBBH}~\cite{Postnov:1998sq,Regimbau:2005tv,Wu:2011ac,Zhu:2011bd,2016PhRvL.116x1103A,Cholis:2016xvo,Chen:2018rzo}.
The latter can come from a number of physical processes, like
quantum fluctuations during inflation~\cite{Salopek:1990re}, the decay of the false vacuum~\cite{Coleman:1977py}, and the evolution of cosmic strings~\cite{Albrecht:1984xv}. 
Note that we discuss different sources separately, and we leave the problem of distinguishing different types of \ac{SGWB} in the future work.

\subsection{Foreground}
There are a great number of \acp{DWD} in our Galaxy, each emitting a quasisinusoidal GW signal. 
Certain frequency bins (especially for lower frequencies) might contain a number of \acp{DWD} emitting GWs with similar amplitudes. 
For these frequency bins, large number of \ac{DWD} signals overlap on top of each other.
In this case, only very strong signals can stand out. 
The signals would then form a foreground by incoherent superposition~\cite{Hils:1990vc,Nelemans:2004qz,Benacquista:2016dwl,Cornish:2017vip}. 

In order to derive the foreground, we first calculate the frequency-domain signals $\wtd{h}(f)$ for all of the \acp{DWD} in each frequency bin. 
The spectrum of initial foreground can be obtained by the sum of squares of $\wtd{h}(f)$ per frequency bin, and the overall spectrum is foreground $S_{\rm DWD}(f)$ plus the instrumental noise $S_{\rm n}(f)$.
We remove the fluctuations in this initial estimate, by binning over the frequencies to use median values within as representatives, which are further downsampled to a handful of points.
A cubic spline fit is later performed on top of these samples~\cite{Cornish:2007if}, and a smooth estimate of the \ac{PSD}, consisting of both the foreground caused by the \acp{DWD} $S_{\rm DWD}(f)$ and the instrumental noise $S_{\rm n}(f)$, is thus obtained.
A source is identified as ``resolvable'' (and later removed from the sample) if the expected \ac{SNR} exceeds the preset detection threshold of 7~\cite{Robson:2017ayy}.
The aforementioned process is then repeated iteratively, continuously lowering the \ac{PSD}, until no new resolvable source is identified.

As the observation time increases, more individual \acp{DWD} can exceed the predetermined threshold of detection.
The identification and later removal of the stronger \acp{DWD} will in turn lower the strength of the foreground.
A preliminary study reveals that for a nominal observation time of 5 years, the accumulated foreground of TianQin would be under the \ac{ASD}~\cite{Huang:2020rjf}.
We fit the foreground with an exponential of the polynomial.
For the numerical coefficients, we direct interested readers to Table II and Fig.~3 in~Ref.~\cite{Huang:2020rjf}.
We are thus confident that for the individual detection and measurement of the other signals at TianQin, the GW foreground has marginal effects.
However, as the foreground has strength comparable to the noise \ac{ASD}, its existence would still overshadow the other types of \ac{SGWB}. 

To the best of the authors knowledge, existing discussions on foreground all based on single GW detector, either LISA~\cite{Robson:2019}, DECIGO~\cite{Kawamura:2011zz}, or TianQin~\cite{Huang:2020rjf}.
Notice that Ref.~\cite{Seto:2020mfd} discussed the detection prospect of \ac{SGWB} with a detector network, but the implication of joint detection on foreground is not explicitly discussed. 
However, notice that the expected operation times of TianQin and LISA might have a certain amount of overlap, and it is meaningful to discuss the foreground under a GW detector network.

For the individually resolvable sources, the optimal \ac{SNR} is defined as the inner product of the expected waveform $\wtd{h}(f)$,
\be \label{eq:InnerProduct}
\rho_{\rm opt}^2 \triangleq \left(h\middle|h\right)=4\Re\int_0^\infty {\rm d}f \frac{\wtd{h}(f)\wtd{h}^*(f)}{P_{\rm n}(f)},
\ee
which denotes the optimal capability for the source with the waveform $\wtd{h}(f)$, and $\Re$ is the real part [which can be ignored here since $h(f)h^*(f)$ is guaranteed to be real].
Unlike in \eq{eq:Pn_Mich} and \eq{eq:Pn_AE}, here $P_{\rm n}$ refers to the \ac{PSD} from the summation of both instrumental noise and foreground.
Besides, the total \ac{SNR} with a GW detector network $\rho_{\rm tot}$ is the root sum square of the \acp{SNR} from each detectors $\rho_i$~\cite{Harms:2008xv,Ma:2017bux},
\be \label{eq:rssSNR}
\rho_{\rm tot} = \sqrt{\sum_i \rho_i^2}.
\ee
Defining the effective \ac{PSD} $P_{\rm n_{tot}}$ with a network of GW detectors~\cite{Wang:2021mou},
\be
\label{eq:Pn_co}
P_{{\rm n}_{\rm tot}}(f)=\frac{1}{\sum_{i}P^{-1}_{{\rm n}_{i}}(f)},
\ee
the total \ac{SNR} can be obtained through \eq{eq:InnerProduct} when an all-sky average on the detector response is assumed, {\it i.e.}, we assume that after averaging, $\wtd{h}(f)$ is kept unchanged across the different detectors,
\bea\label{eq:EffPSD}
\rho_{\rm tot}^2 &=& \sum_i 4\Re\int_0^\infty {\rm d}f \frac{\wtd{h}(f)\wtd{h}^*(f)}{P_{\rm n}^i(f)}\nn \\
&= & 4 \int_0^\infty {\rm d}f \sum_i \frac{\wtd{h}(f)\wtd{h}^*(f)}{P_{\rm n}^i(f)} \nn \\
&= & 4 \int_0^\infty {\rm d}f \frac{\wtd{h}(f)\wtd{h}^*(f)}{P_{{\rm n}_{\rm tot}}(f)}.
\eea

\begin{figure}[t]
	\centering
	\includegraphics[height=6cm]{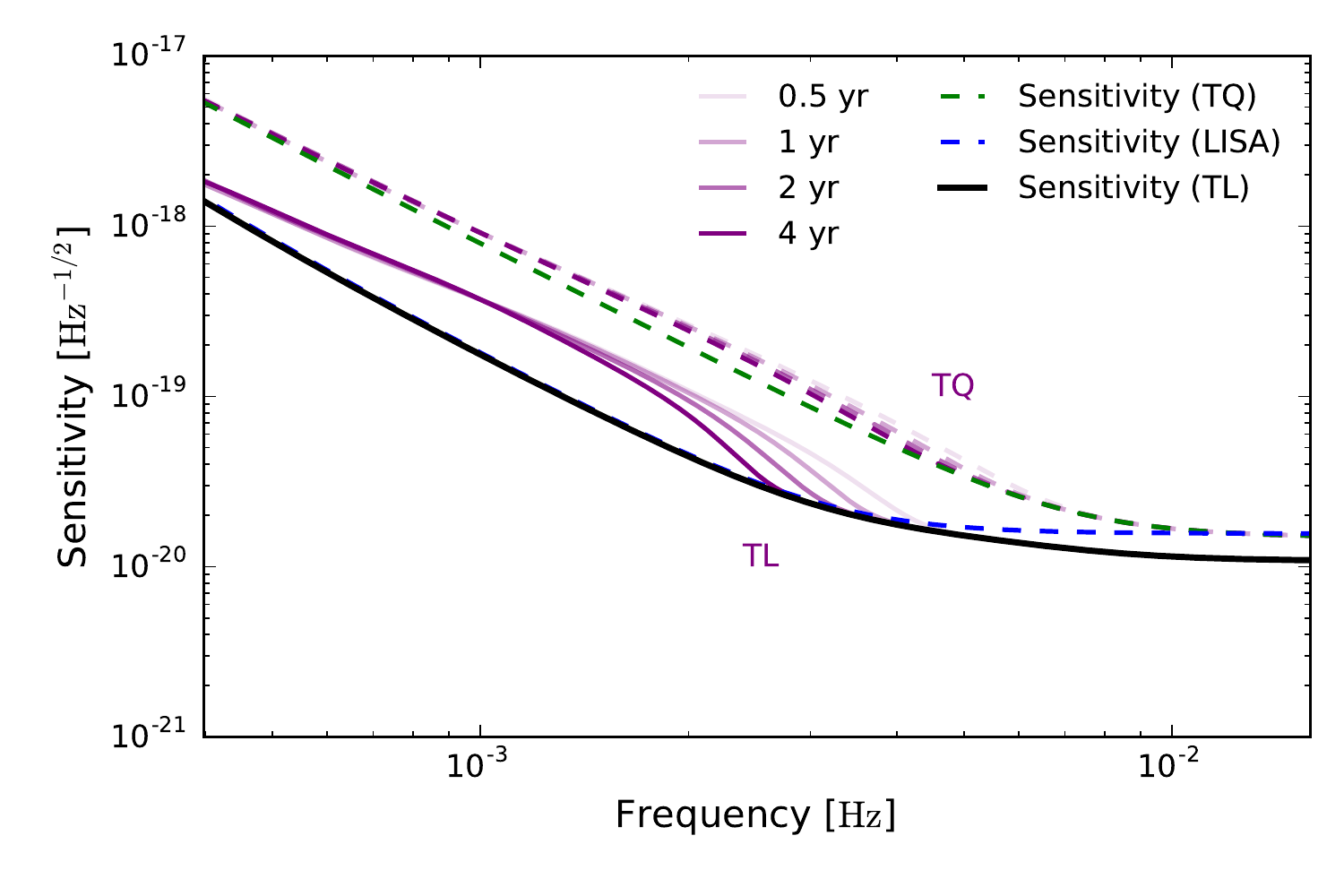}
        \caption{The dashed green curve represents the sensitivity curve of TianQin, and the dashed blue curve represents the sensitivity curve of LISA. The black solid line is the effective sensitivity curve for the joint network of TianQin and LISA. 
        The superposition of foreground and corresponding instrumental noise for the TianQin (dashed line) and TianQin-LISA network (solid line) are shown with different shades of purple, assuming a 0.5, 1, 2, and 4 years operation time, respectively.
        Note that working mode for TianQin has been accounted for in the calculation.}
	\label{fig:Figtl}
\end{figure}

Based on \eqs{eq:rssSNR} and (\ref{eq:Pn_co}), one can estimate the foreground of a detector network by removing the resolvable sources with $\rho_{\rm tot}\ge 7$.
The effective \ac{PSD} is then obtained through smoothing and this process is repeated until no new resolvable source is identified.
In \fig{fig:Figtl}, Combined A with E channels, we convert the \ac{PSD} to the sensitivity curve~\cite{Cornish:2017vip} and present the overall sensitivity that contains both instrumental noise and foreground with 0.5, 1, 2, and 4 years operation time, respectively, assuming a network of TianQin and LISA.
The joint observation of TianQin and the LISA will increase the sensitivity by reducing the effective sensitivity curve, as a result, more sources are expected to be recovered compared with a single constellation, which in return pushes the joint foreground downwards.
From \fig{fig:Figtl} we can easily deduce that the network foreground will be constantly below the sensitivity curve of TianQin.
A network of detectors can increase \ac{SNR} for a given white dwarf binary.
Therefore, with a network, more binaires can be resolved, and one can thus expect the lowering of the joint foreground, as illustrated in the \fig{fig:Figtl}.

\begin{figure}[t]
	\centering
	\includegraphics[height=6cm]{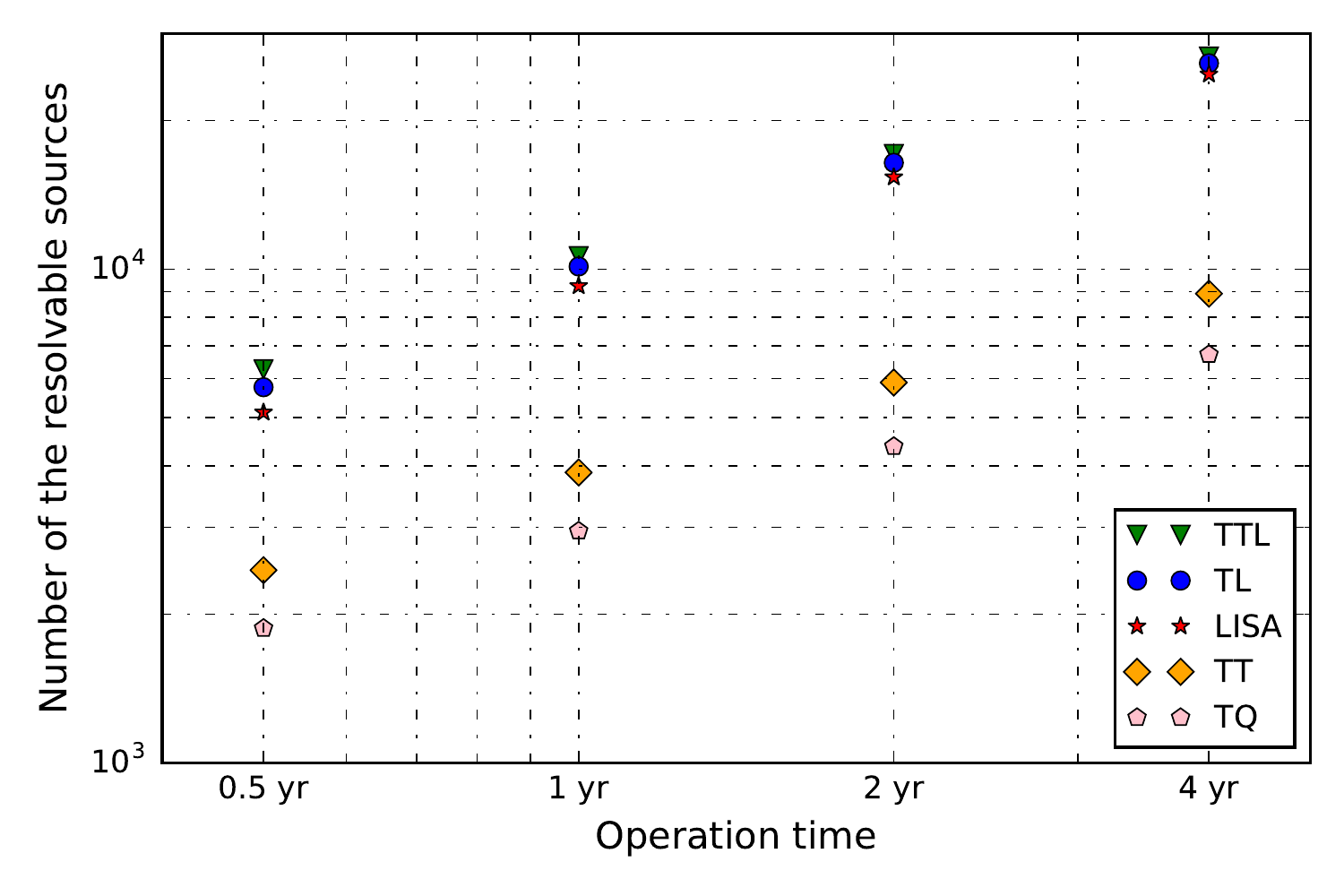}
	\caption{Number of resolvable \acp{DWD} for different detectors/networks. Red stars, blue dots and green triangles represent the union of sources from individual detectors, the TianQin-LISA network, and the network of TianQin I+II and LISA, respectively.}
	\label{fig:n_dwd}
\end{figure}

Moreover, in \fig{fig:n_dwd} we also compare the expected number of resolvable \acp{DWD} with different detectors or detector networks, as a function of different operation time.
Notice that compared with the union of samples from an individual constellation, the network of TianQin and LISA could boost the number of resolvable sources by about 5-12\%, and the number can be further increased to around 8-22\% if TianQin II is included, as shown in \fig{fig:n_dwd}.

\subsection{\ac{SGWB} from compact binary objects}
Astrophysical sources other than the Galactic \acp{DWD} are not expected to be strong enough to reach the level of the noise \ac{PSD}, but the corresponding \ac{SGWB} can still be detectable.

\begin{figure}[t]
	\centering
	\includegraphics[height=6cm]{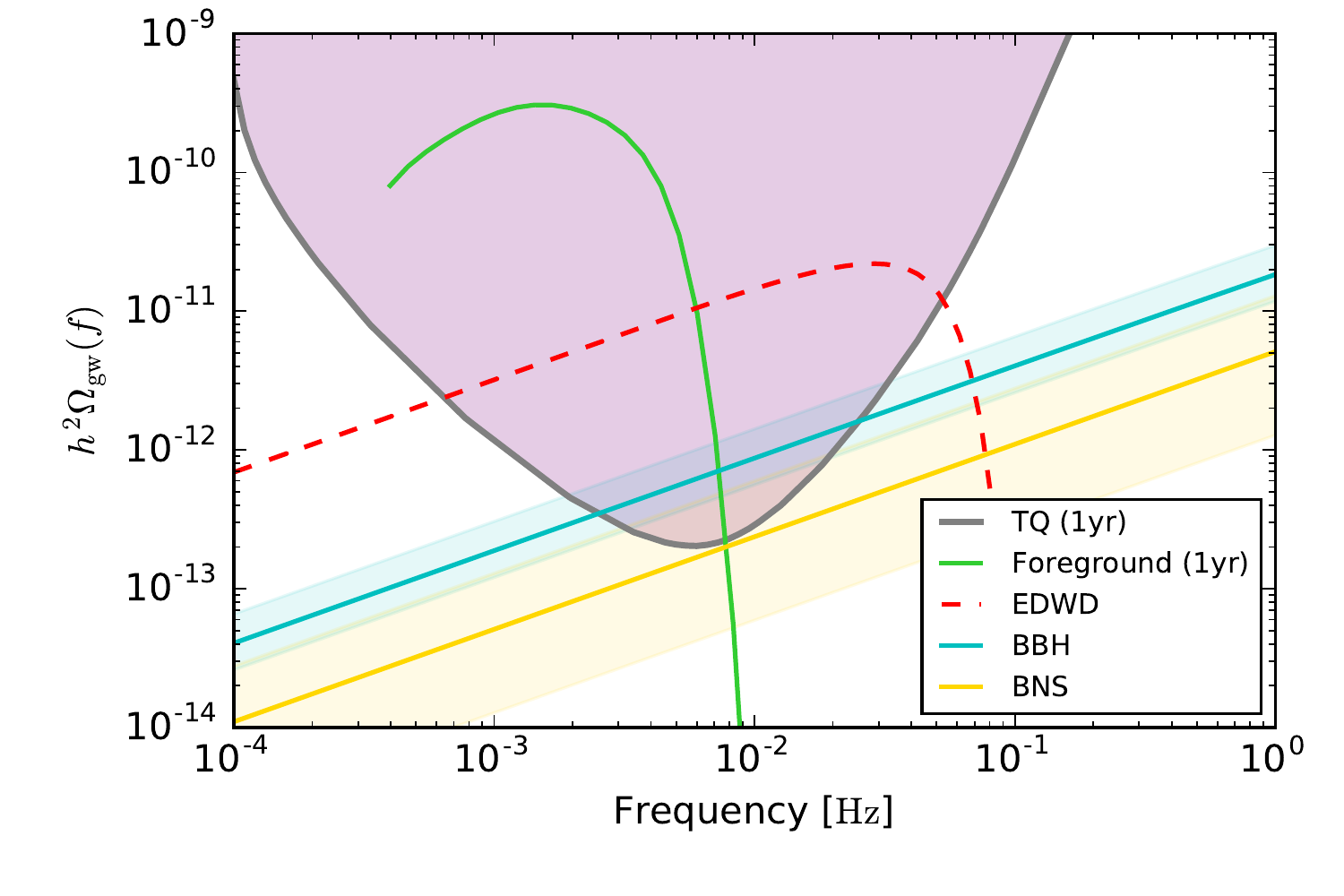}
	\caption{Spectrum density $\Omega_{\rm gw}(f)$ of the Galactic \acp{DWD} (green), the \acp{SBBH} (blue), the \acp{BNS} (orange), and the \ac{EDWD} (red), together with the \ac{PI} sensitivity curve (gray) for TianQin with $1\,\,{\rm yr}$ operation time. The foreground and background produced by the \acp{DWD} are expected to dominate the detection frequency band of TianQin.}
	\label{fig:astro}
\end{figure}

For example, \acp{EDWD} can form such \ac{SGWB}~\cite{Farmer:2003pa,Schneider:2010ks}. 
Following Ref.~\cite{Rosado:2011kv}, we derive the anticipated estimation of the background, as shown in \fig{fig:astro} with the red dashed line. 
We show the anticipated \ac{SNR} in \tab{tab:snr_cbc}.
Of course, due to the lack of direct observation, our conclusions on \ac{EDWD} are subject to uncertainties in the modeling~\cite{Korol:2021pun}.

On the contrary, \acp{CBC} like mergers of \acp{SBBH} as well as \acp{BNS} have been identified by ground-based GW detectors~\cite{LIGOScientific:2018mvr,Abbott:2020uma,LIGOScientific:2020stg,Abbott:2020khf,Abbott:2020tfl}.
Such compact binaries can emit $\rm{mHz}$ GW signals during the early inspiral stage months to years prior to the merger, which can be detected by space-borne GW detectors \cite{Liu:2020eko,Sesana:2016,Hu:2017yoc}.
It is expected that such signals can stack up to form a \ac{SGWB}.

With the $k$th type of the compact binaries described by the population properties $\boldsymbol{\theta}_{k}$ (like the component masses, spin, orbital eccentricity, and progenitor metallicity, {etc.}), following Ref.~\cite{TheLIGOScientific:2016wyq,Abbott:2017xzg}, the spectrum density of \ac{SGWB} $\Omega_{\rm gw}(f)$ from each component can be formulated as the superposition of each binary's energy spectrum:
\be
\label{eq:omegastro}
\Omega_{\rm gw,0}(f)=\frac{f}{\rho_{\rm c}}
\int\,{\rm d}\boldsymbol{\theta}_{k}\int_{0}^{z_{\rm max}}\,{\rm d}z
\frac{R_{\rm m}(z,\boldsymbol{\theta}_{k})\frac{{\rm d}E_{\rm gw}}{{\rm d}f_{\rm s}}(f_{\rm s},\boldsymbol{\theta}_{k})}{(1+z)H(z)},
\ee
where $R_{\rm m}(z,\boldsymbol{\theta}_{k})$ is the merger rate, and $\frac{{\rm d}E_{\rm gw}}{{\rm d}f_{\rm s}(f_{\rm s},\boldsymbol{\theta}_{k})}$ profiles the energy spectrum emitted in the source frame $f_{\rm s}=f(1+z)$. 
$H(z)$ is the Hubble parameter, and by making the maximum redshift $z_{\rm max}=10$ (above which not many astrophysical black holes are expected to form due to the lack of star formation), $H(z)$ can be safely approximated as $H_{0}\sqrt{\Omega_{\rm m}(1+z)^{3}+\Omega_{\Lambda}}$ ~\cite{Aghanim:2018eyx}. 
In this work, we do not consider the influence of the orbital eccentricity for the binaries. 
Furthermore, instead of the evolving merger rate density model~\cite{Callister:2016ewt,Dvorkin:2016wac,Nakazato:2016nkj,Inayoshi:2016hco}, we adopt the official merger rate derived from observations of the ground-based detectors, in which the redshift evolution is also not included~\cite{Abbott:2020gyp}, {i.e.}, $R_{\rm m}(z,\boldsymbol{\theta}_{k})$ is independent of the redshift. 
For the mass distribution, we adopt the ``POWER LAW + PEAK'' model in~Ref.~\cite{Abbott:2020gyp} for the stellar-mass \acp{SBBH}, which is equivalent to ``Model C'' described in~Ref.~\cite{LIGOScientific:2018jsj}, where $5M_{\odot}\le m_{2}\le m_{1}\le 100M_{\odot}$. 
As for the \acp{BNS}, we adopt a uniform component-mass distribution~\cite{TheLIGOScientific:2016pea} ranging from 1-$2.5 M_{\odot}$ without redshift evolution. 
We adopt the $R_{\rm m}(0,\theta_{\rm BBH})$ value $23.9^{+14.9}_{-8.6}\,{\rm{Gpc^{-3}\,\,yr^{-1}}}$ and the $R_{\rm m}(0,\theta_{\rm BNS})$ value $320^{+490}_{-240}\,\,{\rm{Gpc^{-3}\,yr^{-1}}}$ provided by~Ref.~\cite{Abbott:2020gyp} as the merger rate.
For the energy spectrum, ${\rm d}E_{\rm gw}/{\rm d}f_{\rm s}$, we adopt the same format as~Ref.~\cite{Perigois:2020ymr}, which considers the updated IMR phenomenological waveform~\cite{Ajith:2009bn}.
The spectrum densities of the two types of \acp{CBC} are shown in \fig{fig:astro}, which are expected to be submerged by the \acp{DWD}'s in the frequency band of TianQin.
In \tab{tab:snr_cbc} we list the expected \acp{SNR} for \acp{SBBH} and \acp{BNS}, considering the above range of merge rate $R_{\rm m}$ and assuming 1 year operation time. Compared with \acp{DWD}, the \acp{SNR} of the \acp{SBBH} and \acp{BNS} are around 1 order of magnitude lower.
It has also been proposed that \acp{SBBH} can originate from \acp{PBH}~(e.g., Refs.~\cite{Mandic:2016lcn,Clesse:2016ajp}).
Since the spectrum is not expected to be significantly different from that of \acp{SBBH} of astrophysical origin, we do not discuss it as a separate case.

\begin{table}
	\begin{center}
		\caption{\acp{SNR} for the \acp{SGWB} from \acp{SBBH}, \acp{BNS} and \acp{EDWD}, assuming 1 year operation time.}\label{tab:snr_cbc}
		\setlength{\tabcolsep}{3mm}
		\renewcommand\arraystretch{1.5}
		\begin{tabular}{*{5}{|c}|}
			\hline
			& SBBH                & BNS                 & EDWD              \\
			\hline
			TQ                    & \multirow{2}*{$3.3^{+2.0}_{-1.2}$} &  \multirow{2}*{$0.9^{+1.3}_{-0.7}$}    &\multirow{2}*{$50$}   \\
			(null channel)        &                     &            &  \\
			\hline 
			TQ I+II               & \multirow{2}*{$1.3^{+0.8}_{-0.5}$}&  \multirow{2}*{$0.4^{+0.5}_{-0.3}$}    &\multirow{2}*{$20$}   \\
			(cross correlation)   &                     &            &  \\
			\hline
			TQ+LISA               & \multirow{2}*{$1.0^{+0.6}_{-0.4}$}   &  \multirow{2}*{$0.3^{+0.4}_{-0.2}$}    &\multirow{2}*{$14$}   \\
			(cross correlation)   &                     &            &  \\
			\hline  
		\end{tabular}
	\end{center}
\end{table}

A similar procedure is also applied to the \acp{MBHB} with masses between $10^{2}$-$10^{8}M_{\odot}$ to estimate the corresponding \ac{SGWB}.
\acp{MBHB} can emit strong GW signals in a wide frequency range, and the Pulsar Timing Array is expected to resolve the individual mergers as well as detect the \ac{SGWB} in the $\rm{nHz}$ band~\cite{Kelley:2017lek,Dvorkin:2017vvm,Arzoumanian:2020vkk}.
Space-borne GW detectors operating in the $\rm{mHz}$ band can observe the individual mergers~\cite{Wang:2019,Feng:2019wgq}, but the corresponding \ac{SGWB} is unlikely to be detectable~\cite{Sesana:2004gf,Sesana:2007sh,Sesana:2008mz}. 
Based on the astrophysical models for \ac{MBHB} mergers adopted by~Ref.~\cite{Wang:2019}, we conclude that the \ac{SGWB} is indeed too weak to be detected by TianQin.

We also consider sources with other origins, like core-collapse supernovae~\cite{Ferrari:1998ut,Crocker:2017agi}, rotating neutron stars~\cite{Owen:1998xg,Rosado:2012bk}, and magnetars~\cite{Marassi:2010wj,Cheng:2017kmv}.
We have studied their spectra and concluded that their strengths are not strong enough to be detected by TianQin.

%%%%%%%%%%%%%%%%%%%%%%%
\subsection{\ac{SGWB} from inflation}
%%%%%%%%%%%%%%%%%%%%%%%

In addition to the astrophysical-origin backgrounds, the \ac{SGWB} can be also generated by the cosmological process. 
During the inflation, quantum tensor fluctuations of the spacetime can be stretched beyond the horizon. This process can amplify the fluctuation. When the Universe slows down, they will reenter the horizon and form the primordial GWs~\cite{1976JETPL..23..293G,1979JETPL..30..682S,Polarski:1995jg,2016NCimR..39..399G}. 
This constitutes an irreducible source of the GWs that are expected from any inflationary model~\cite{2000ASSL..247.....M}.
For the tensor modes with frequencies of a few $\rm mHz$, they reentered the horizon in the deep radiation-dominated era.
The propagation of this primordial background until the present leads to the GW energy density spectrum today~\cite{Maggiore:2018sht}:

\be
h_{0}^{2} \Omega_{\rm gw}(f)=
0.016\cdot h_{0}^{2} \Omega^0_{r}\,r\,\mathcal{P}_{\mathcal{R}}
\bigg(\frac{f}{f_{\rm{CMB}}}\bigg)^{n_{\rm T}},f\ge10^{-4}\,\,{\rm{Hz}},
\ee
where the Hubble constant $h_{0}=0.674$~\cite{Aghanim:2018eyx}, the radiation energy density today $\Omega^0_{r}=h_{0}^{-2}\cdot4.17\times10^{-5}$~\cite{Hu:2001bc}, and the primordial curvature power spectrum $\mathcal{P}_{\mathcal{R}}\simeq 2.14\times10^{-9}$ is associated with the pivot frequency $f_{\rm{CMB}}=7.73\times10^{-17}\,\,{\rm{Hz}}$~\cite{Akrami:2018odb}.

Clearly, the order of $h_{0}^{2} \Omega_{\rm gw}$  depends on the tilt of the tensor spectrum $n_{\rm T}$ and the tensor-to-scalar ratio $r$~\cite{Caprini:2018mtu}. For the minimal model where a single scalar field drives a slow-roll inflation, these two parameters obey the consistency relation $n_{\rm T} = -r/8$. As a result, the prediction for $h_{0}^{2} \Omega_{\rm gw}$ in the minimal inflation model is entirely determined by the size of $r$. The most stringent bound $r \lesssim 0.056$ from {\it Plank}~\cite{Akrami:2018odb} demands a very small $|n_{\rm T}|$, resulting in a nearly scale-invariant power spectrum with $h_{0}^{2} \Omega_{\rm gw} (f) \lesssim 6\times 10^{-17}$. This value is at least 3 orders of magnitude below the TianQin sensitivity curve and the primordial background is thus inaccessible to TianQin.

\begin{table*}[t]
	\begin{center}
		\caption{Ingredients of the GW spectra from the \ac{PT}, where $\alpha_{\infty}$ determines the weakest transition above which a portion of the vacuum energy converts into the kinetic energy driving the bubble expansion~\cite{Ellis:2019oqb}. The Hubble parameter at the moment of GW production with the total number of relativistic degrees of freedom $g_{*}$ and temperate $T_{*}$ at the time when GWs are produced is $h_{*}=1.65\times10^{-5}{\rm~Hz}
\bigg(\frac{T_{*}}{100 {\rm~GeV}}\bigg)\bigg(\frac{g_{*}}{100 {\rm~GeV}}\bigg)^{1/6}$.}
\label{tab:para_pt}
		\setlength{\tabcolsep}{3mm}
		\renewcommand\arraystretch{2}
		{
			\begin{tabular}{|c|c|c|c|c|c|c|}
				\hline
				$i$         & $p$  & $q$  & $\tilde{\Delta}_{i*} (v_w)$  & $\frac{\tilde{f}_{i*}}{\beta}$ 
				&  $S_{i}(f,\tilde{f}_{i})$ &  $\kappa_{i}(\alpha)$       \cr 
				\hline
				col       & 2  & 2 & $\frac{0.44v_{w}^{3}}{1+8.28v_{w}^{3}}$  & $\frac{0.31}{1-0.051v_{w}+0.88v_{w}^{2}}$ 
				&  $\frac{3.8(f/\tilde{f}_{i})^{2.9}}{1+2.9(f/\tilde{f}_{i})^{3.8}}$ &  ${\rm max}\big[1-\frac{\alpha_{\infty}}{\alpha},0\big]$     \cr
				\hline
				\multirow{1}{*}{sw} & 1  & 2 & $0.157v_{w}H_{*}\tau_{\rm sw}$  & $1.16\frac{L_{\rm s}}{R_{*}}$ &   $\left(\frac{f}{\tilde{f}_{i}}\right)^{3}\left(\frac{7}{4+3(f/\tilde{f}_{i})^{2}}\right)^{7/2}$    &  
				\multirow{2}{*}{$\left\{\begin{gathered}
					\kappa(\alpha_{N})|_{\alpha_{N}=\alpha}, \alpha\le\alpha_{\infty}\\
					\frac{\alpha_{\infty}}{\alpha}\kappa(\alpha_{N})|_{\alpha_{N}=\alpha_{\infty}},\alpha>\alpha_{\infty}
					\end{gathered}
					\right.$}  \cr
				\cline{1-6}
				\multirow{1}{*}{turb}       & 1  & $\frac{3}{2}$ & $20v_{w}(1-H_{*}\tau_{\rm sw})$  & $1.33\frac{L_{\rm s}}{R_{*}}$
				& $\frac{(f/\tilde{f}_{i})^{3}}{[1+(f/\tilde{f}_{i})]^{\frac{11}{3}}(1+8\pi f/h_{*})}$  &    \cr
				\hline
			\end{tabular}}
	\end{center}
\end{table*}

Beyond irreducible emission, GW backgrounds with high amplitude and a significant deviation from scale invariance, can also be produced if new species or symmetries are introduced during inflation~\cite{Baumann:2007zm,Cheung:2007st,Barnaby:2010vf,Kobayashi:2010cm,Senatore:2011sp,Biagetti:2013kwa}. However, the discussion of these inflationary backgrounds is beyond the scope of this paper. 

\subsection{\ac{SGWB} from first-order \ac{PT}}
In addition to the primordial GWs formed during the inflationary period, \ac{SGWB} can also be generated after inflation where the important sources include the induced GWs in the \ac{RD} era~\cite{Saito:2008jc,Cai:2018dig,Bartolo:2018evs}, nonperturbative effects in the reheating epoch~\cite{Dufaux:2007pt}, of the early Universe~\cite{Hogan:1986qda}, and cosmic defects~\cite{Vachaspati:1984gt,Krauss:1991qu}. 
In this section we discuss the last two sources as they have a strong relation with the inflationary models and TeV-scale physics.

%This is typically driven by some scalar field(s) acquiring a non-zero vacuum expectation value within a certain vacuum manifold M.
Generating the \ac{SGWB} from the first-order \ac{PT}s~\cite{Kamionkowski:1993fg,Grojean:2006bp,Caprini:2015zlo,Mazumdar:2018dfl,Di:2020ivg} is a theoretically favorable scenario given the fact that the electroweak symmetry of elementary particle physics is broken in the present Universe. 
A first-order \ac{PT} in the early Universe corresponds to a tunneling process from a symmetric phase to a broken phase with a lower energy (true vacuum) through the nucleation of bubbles of the true vacuum that receive the vacuum energy released from the PT.
%The vacuum bubbles of the scalar field nucleated during the \ac{PT}  
%and become an important source of producing the GWs. 
Following the nucleation~\cite{Hogan:1984hx}, the bubbles expand and eventually collide with others, eventually causing gravitational radiation due to the destruction of the spherical symmetry preserved in a single vacuum bubble~\cite{Kosowsky:1992rz}.
%\JY{a stochastic background of GWs would be generated during the process of bubble expansion and collision following the bubble nucleation with typical redshifted spectrum frequency around $\mathcal{O}(10^{−4} - 10^{−2}) {\rm Hz}$, }
This physical process has been described by the envelope approximation model~\cite{Kosowsky:1992rz} which assumes that the liberated vacuum energy is entirely concentrated in the bubble wall and neglects the overlapping regions of colliding bubbles.
% and assumes the contribution to gravitational radiation only arises from the uncollided envelop where the vacuum energy is stored in these regions.}
 
A recent successful large-scale hydrodynamical simulation~\cite{Hindmarsh:2013xza} has revealed that, for a PT happening in the thermal cosmic medium, the majority of the vacuum energy stored in the bubble wall does not dissipate after the collision but rather is transferred to the surrounding plasma fluid, which may give rise to a substantial amount of the GW radiation in the form of sound waves and turbulence due to the movement of the plasma fluid caused by the bubble collision~\cite{Hindmarsh:2015qta}. Nonetheless, our understanding to either the acoustic production mechanism or the origin of the vortical turbulence is still not clear. The sound-shell model~\cite{Hindmarsh:2016lnk} has attempted to explain the former puzzle. 
Ignoring the interference between one source and another, we linearly superimpose these three contributions in computing the total \ac{SGWB} observed today,
\be
\label{eq19}
h_0^{2}\Omega_{{\rm gw}, 0}(f)=\sum_{i} h_0^{2}\Omega_{i,0}(f),
\ee
where the sum includes the collision occurring during the \ac{PT}, and acoustic and turbulent phases after the \ac{PT}. 
The individual contributions after taking into account the redshift can be written in a compact form,
\bea
\nn
h_0^{2}\Omega_{i,0}(f)&=&1.67\times10^{-5}\left(\frac{100}{g_{*}(T_{*})}\right)^{1/3} S_{i}(f,\tilde{f}_{i}) \\
&&\left(\frac{H_{*}}{\beta}\right)^{p}
\left(\kappa_{i}(\alpha ) \frac{\alpha}{1+\alpha}\right)^{q}\tilde{\Delta}_{i*}(v_w),
\label{eq:omegapt}
\eea
where the indices $p,q$, the spectral function $S_{i}(f,\tilde{f}_{i})$, the peak amplitude $\tilde{\Delta}_{i*} (v_w)$, and the efficiency factor $\kappa_{i}(\alpha)$ are collected in \tab{tab:para_pt}. 
%\JY{can add some comments on the enhancement factor the inverse \ac{PT} duration $\beta^{-1}$, relative importance and etc. such as } 
Similarly, the redshifted peak frequency today for each source takes the form~\cite{Jinno:2016vai},
\be
\tilde{f}_{i}=1.65\times10^{-5}{\rm~Hz}\frac{\tilde{f}_{i*}}{\beta}
\left(\frac{\beta}{H_{*}}\right)\left(\frac{T_{*}}{100 {\rm~GeV}}\right)\left(\frac{g_{*}(T_{*})}{100}\right)^{\frac{1}{6}},
\ee
with the peak frequency before the redshift $\tilde{f}_{i*}$ given in \tab{tab:para_pt}. 

Two important remarks on the peak parameters are placed in order. 
First, a recent study introduces the sound shell thickness $L_{\rm S}=R_{*}(1-c_{\rm s}/v_{w})$~\cite{Hindmarsh:2017gnf} in addition to the mean bubble separation $R_{*}=(8\pi)^{1/3}v_{w}/\beta$~\cite{Enqvist:1991xw}, with $c_{\rm s}=1/\sqrt{3}$ being the the sound speed. This modification respects the new observation that the length scale of the acoustic waves is characterized by the thickness of the sound shell and will cause a redshift on the peak frequency of the GW power spectrum as the bubble velocity $v_w$ decreases.
The other improvement made in~\cite{Ellis:2019oqb} involves an accurate estimate on the duration of the acoustic production,
\bea
\tau_{\rm sw}&=&\min \left[{H^{-1}_{*}},R_{*}/U_{\rm f} \right],
\eea
where $U_{\rm f}$ is the average square root of the fluid velocity~\cite{Espinosa:2010hh}. 
This treatment is particularly crucial for the strong \ac{PT} where the fluid develops into turbulence within one Hubble time~\cite{Wang:2020jrd}; otherwise, the contribution from the turbulent phase would be underestimated. 

\begin{figure}[t]
	\centering
	\includegraphics[height=6cm]{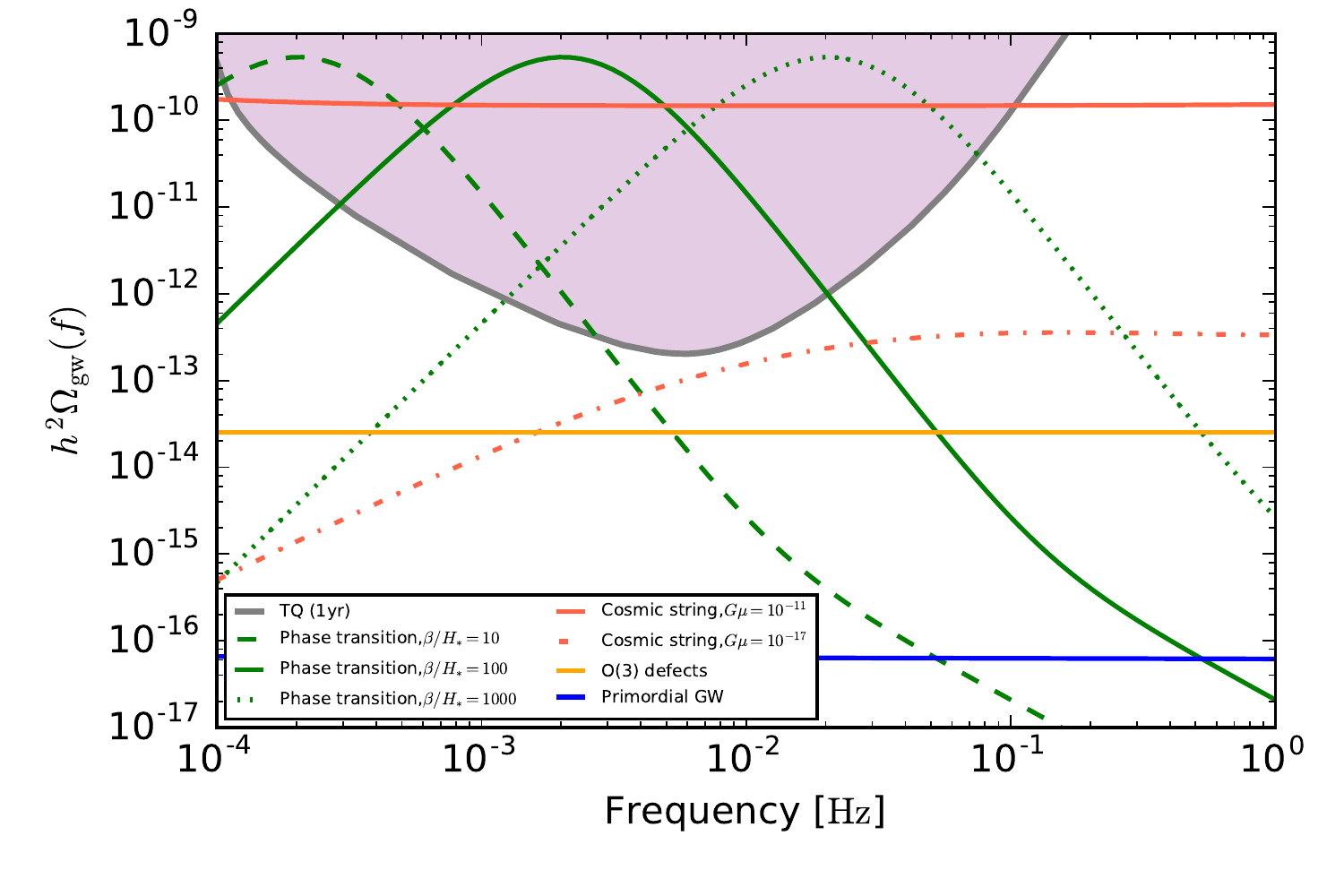}
	\caption{To obtain the green line, we neglect the contribution of the scalar field, and set $\alpha=0.5$, $v_{w}=0.95$, $\beta/H_{*}$=10, 100, 1000, and $T_{*}=100\,{\rm{GeV}}$.
		As for the \ac{SGWB} produced by inflation, we set the tensor-to-scalar ratio $r=0.056$~\cite{Akrami:2018odb}. 
		As for the cosmic defects, we set $G\mu=10^{-11}$, $10^{-17}$ for string loops decay, and $N=3$, $v=10^{16}\,\,{\rm{GeV}}$~\cite{Ade:2013xla} for $O(N)$ theory to obtain the spectrum, respectively.
		We cannot determine the exact value but a certain range of the \ac{SNR} in \tab{tab:snr}.}
	\label{fig:cos}
\end{figure}

From \eq{eq:omegapt} it is apparent that the redshifted \ac{SGWB} spectrum generated from the PT is entirely determined by the properties of the PT including the PT scale $T_*$, the PT strength $\alpha$, and the PT duration (normalized to the Hubble parameter) $\beta/H$.
As an example, we consider a PT occurring at the electroweak scale $T_*=100\,\,{\rm{GeV}}$ with a moderate strength $\alpha=0.5 \le\alpha_{\infty}$ and the short duration $\beta/H=10,100,1000$ and adopt the quantitative estimate performed in~Ref.~\cite{Hindmarsh:2017gnf}.
%In this case, the bubble collision the vacuum energy stored in the nucleated bubbles of the scalar field is negligible. The GW from
The produced redshifted \ac{SGWB} spectrum shown in \fig{fig:cos} is typically peaked within the frequency band of $\mathcal{O}$($10^{-4}$--$10^{-1}$) Hz, where the space-borne GW detectors (including TianQin) reach the best sensitivity. 
This could lead to a substantially larger \ac{SNR} for the detectable \ac{SGWB}. 
The corresponding \acp{SNR} are shown in \tab{tab:snr_pt}.

\begin{table}
	\begin{center}
		\caption{\acp{SNR} for the \ac{SGWB} from a \ac{PT}, where the parameters are fixed to $\alpha=0.5$, $v_{w}=0.95$, $T_{*}=100\,\,{\rm{GeV}}$, and we set $\beta/H_{*}=10,100,1000$ and assume 1 year operation time.}
		\label{tab:snr_pt}
		\setlength{\tabcolsep}{1mm}
		\renewcommand\arraystretch{1.5}
		\begin{tabular}{*{5}{|c}|}
			\hline
		                          & $\beta/H_{\ast}=10$              & $\beta/H_{\ast}=100$  
		                          & $\beta/H_{\ast}=1000$               \\
			\hline
			TQ                    & \multirow{2}*{$22$} &  \multirow{2}*{$320$}    &\multirow{2}*{$390$}   \\
			(null channel)        &                     &            &  \\
			\hline 
			TQ I+II               & \multirow{2}*{$9.0$}&  \multirow{2}*{$160$}    &\multirow{2}*{$170$}   \\
			(cross correlation)   &                     &            &  \\
			\hline
			TQ+LISA               & \multirow{2}*{$43$}   &  \multirow{2}*{$190$}    &\multirow{2}*{$36$}   \\
			(cross correlation)   &                     &            &  \\
			\hline  
		\end{tabular}
	\end{center}
\end{table}

In addition to a single-peak spectrum, more exotic shapes of the GW spectrum such as the multipeak one are also possible~\cite{Huang:2017laj}. 
%We postpone for future work a delicate analysis and the detectability of this background at TianQin.
Since a first-order \ac{EWPT} is not expected in the SM~\cite{Kajantie:1995kf,Kajantie:1996mn}, any detection of the GWs from the \ac{EWPT} would provide a unique probe for the nature of the \ac{EWPT} triggered by new physics beyond the SM that is difficult to be measured at colliders.

First-order \ac{EWPT} is not expected in the SM~\cite{Kajantie:1995kf,Kajantie:1996mn}. 
Therefore, GW detection of \ac{SGWB} from the \ac{EWPT} can serve as a unique probe for the new physics beyond the SM.

%%%%%%%%%%
\subsection{\ac{SGWB} from cosmic defects}
The formation of cosmic defects is another important product of a \ac{PT}, particularly for the one that happens in the inflationary period or shortly after inflation. 
Such formation is closely associated with a (spontaneous) breaking of continuous symmetry and typically arises in the case that the vacuum produced from the \ac{PT} is topologically nontrivial~\cite{Vilenkin:2000jqa}.
Depending on the dimensionality of the defect network, the cosmic defects are divided into topological and nontopological defects. The topological defects include domain walls, strings, and monopoles~\cite{Vilenkin:1981kz}, which will be specifically discussed below. 

In what follows we first describe the irreducible emission of GWs from any type of cosmic defect network. 
Once formed during the \ac{RD} era, any defect network always radiates GWs with an exact scale-invariant energy density spectrum~\cite{Figueroa:2012kw}. 
This remarkable feature of scale-invariance was first confirmed by numerical simulation~\cite{Giblin:2011yh}.
Nonetheless, the \ac{SGWB} spectrum produced in the full cosmic history is not scale invariant~\cite{Figueroa:2020lvo}.
This is due to a substantial GW emission from the modes that entered during the \ac{MD} 
over the one from the \ac{RD} era below the frequency at matter-radiation equality. 
In contrast, this contribution is extremely weak in the TianQin frequency band and thus does not destroy the scale invariance in the GW spectrum.
For simplicity, we consider the irreducible GW background during the \ac{RD} era. 
In this case, their amplitude is only characterized by the symmetry-breaking scale and the nature of the defects.

The formation of cosmic defects is possible when a global $O(N)$ symmetry spontaneously breaks into $O(N-1)$ symmetry. 
In particular, nontopological defects such as the texture agree with the large-$N$ limit $N\geq 3$. 
The \ac{SGWB} spectrum today corresponding to these sources takes the universal form \cite{Fenu:2009qf,Figueroa:2012kw}
\be
\label{eq:CSomega}
h^{2}_0\Omega_{{\rm gw},0}\simeq\frac{650}{N}\tilde{f}(N) h^{2}_0\Omega^0_{\rm r}\left(\frac{\upsilon}{M_{\rm Pl}}\right)^{4},
\ee
where the prefactor $\tilde{f}(N)$ is included to match the numerical result and, to a good approximation, can be parametrized as $\tilde{f}(N)=1.1+45/N^{2}$~\cite{Figueroa:2012kw}. Therefore, \eq{eq:CSomega} suggests that the amplitude of this \ac{SGWB} today is suppressed by $N$ and increases with the factor $(v/M_{\rm Pl} )^4$, with $v$ being the scale of the symmetry breaking 
%(the VeV acquired by the scalar embedded in the $N$-component field) 
and the Planck mass $M_{\rm Pl}\simeq 1.22\times10^{19} {\rm~GeV}$. 
We find that the cosmic defects from an $O(3)$ breaking at the scale $v\lesssim 10^{16} \rm{~GeV}$ produce a GW signal whose magnitude is roughly 1 order below the TianQin sensitivity curve. This indicates that only the cosmic defects that formed close to the Planck scale are detected by TianQin. 

As a special example, we focus on the one-dimensional topological defect: cosmic strings~\cite{Caldwell:1991jj,Sanidas:2012ee}.
In general, cosmic strings can be either fundamental (super) strings correspnding to the string theories or stable strings employed in the field theories~\cite{1986qgc..conf..269V}, {\it i.e.} the structurally simplest $O(2)$ or $U(1)$ theories~\cite{Hindmarsh:1994re,Turok:1991qq,Durrer:2001cg}.
For an infinite thin string [\ac{NG} string], either type of strings is typically characterized by a single dimensionless quantity $G \mu$ (with Newton's constant $G$ and the string energy density $\mu$), which is related to the symmetry-breaking scale $v$ via $G\mu = \pi (v/M_{\rm Pl})^2$~\cite{Bethke:2013aba}.
However, these two types of strings have different properties, such as how the strings will collide, or how often it occurs~\cite{Auclair:2019wcv}.
This main difference significantly affects the evolution of the string network and the production of GWs expected from the decay of the loops.

The history of a string network from its formation to the present time is described as follows. 
Once created, a string network is stretched by the cosmological expansion, and the relativistic motion of the strings leads to string self-intersects or two curved strings collide, this process can produce loops. 
Loops smaller than the horizon decouple from the cosmological evolution and oscillate under their own tension, slowly decaying into GWs, whose spectrum is primarily constituted by bursts from cusps, kinks, and kink-kink collisions\footnote{Current observation from pulsar timing array puts limit on such process, so if it happens, it happens rarely~\cite{Arzoumanian:2018saf}.}. 

Other than the subhorizon loops, the historical remnants of a network also contain superhorizon loops and even open strings, both of which stretch across a Hubble volume. 
Although these strings also emit GWs coming from the accumulation of their small-scale structure, for the \ac{NG} cosmic string networks this background plays an insignificant role in the emission of the GW background.
Thus, we will not include this contribution in the following analysis.

Recent simulations of the cosmic string networks~\cite{Ringeval:2017eww} favor the large subhorizon loops with sizes $\alpha_{l}\gg \Gamma G\mu$ produced by the network. In this case, the total \ac{SGWB} spectrum from the whole network of the loops can be approximately estimated by means of adding the GW emission from all of the loops throughout the entire history of the Universe.
This includes the contributions of the loops that are created and decay in the \ac{RD} era, the \ac{RD} era loops that decay during the \ac{MD} era, and the loops created in the \ac{MD} era: 
\be
\Omega_{\mathrm{gw}}(f)=\Omega_{\mathrm{gw}}^{\rm r}(f)+\Omega_{\mathrm{gw}}^{\rm r m}(f)+\Omega_{\mathrm{gw}}^{\rm m}(f).
\ee
Each component takes the common form~\cite{Blanco-Pillado:2017oxo,Auclair:2019wcv} 
\be
\label{eq:omegaCS}
\Omega^i_{\rm gw,0}(f)=\frac{8\pi (G\mu)^{2}}{3H_{0}^{2}f}\sum_{k=1}^{\infty}
2k {\bf P}_k \!\int\!{\rm d}z\frac{{\bf n}_i\left({2k\over (1+z)f},t(z)\right)}{H(z)(1+z)^{6}},
\ee
where $H(z)\simeq H_0 \sqrt{\Omega_{\rm m}(1+z)^3+\Omega_{\rm r} (1+z)^4}$ for the matter-plus-radiation Universe. 
Evidently, the amplitude of the \ac{SGWB} spectrum is strongly determined by the averaged loop power spectrum ${\bf P}_k$ emitted in each mode $k$ for a particular loop. 
We adopt the BOS spectrum~\cite{Blanco-Pillado:2017oxo} with the cusp events dominated on the smooth loops oscillating in the large-$k$ mode. 
For the loops containing cusps, kinks, and kink-kink collisions, ${\bf P}_k$ scales as $k^{-4/3}$~\cite{Vachaspati:1984gt}, $k^{-5/3}$~\cite{Allen:1991bk}, and $k^{-2}$~\cite{Vilenkin:2000jqa}, respectively. 
Furthermore, an accurate average power spectrum of loops can be yielded after including the gravitational backreaction.
The other key ingredient in \eq{eq:omegaCS} involves the loop number density ${\bf n}(\alpha_{l},t)$ of the non-self-intersecting, subhorizon loops of size $\alpha_{l}$ at cosmic time $t(z)$.
Based on numerical simulation~\cite{Blanco-Pillado:2013qja}, the loop distribution reads \cite{Sousa:2013aaa}
\bea
{\bf n}(\alpha_{l},t)=\left\{
\begin{array}{lr}
\frac{0.18}{t^{4}(\alpha_{l}+\Gamma G\mu)^{5/2}} \Theta(0.1-\alpha_{l}),\,\, ({\rm RD}, z\ge z_{\rm eq}) \\
\\
\frac{0.18(2 \sqrt{\Omega^0_{\rm r}})^{3/2}}{t^{5/2}(\alpha_{l}+\Gamma G\mu)^{5/2}}(1+z)^{3},z\le z_{\rm eq}\\
\\
\frac{0.27-0.45\alpha_{l}^{0.31}}{t^{4}(\alpha_{l}+\Gamma G\mu)^{2}} \Theta(0.18-\alpha_{l}),\,\, ({\rm MD}, z\le z_{\rm eq}),
\end{array}
\right.
\eea
where $z_{\rm eq}$ corresponds to the redshift at  matter-radiation equality~\cite{Abbott:2017mem}.

As can be seen, the resulting spectrum of the GWs emitted by a network of cosmic strings is sensitive to the various properties of the string network, such as (i) the string tension $G\mu$, (ii) the size of cosmic string loops relative to the horizon at birth $\alpha_{l}$, 
and (iii) the cutoff for $k_*$ in the numerical procedure.

In \fig{fig:cos} we show the \acp{SGWB} from the above cosmological sources. 
We show that TianQin will be sensitive to string tensions with $G \mu \gtrsim 10^{-17}$ for strings, where the upper limit of $G \mu$ is $10^{-11}$~\cite{Blanco-Pillado:2017oxo,Ringeval:2017eww}. 
In the high frequency, the \ac{SGWB} of the \ac{RD} cosmic string may appear oblique rather than flat. The exact form is model dependent. Therefore, the \ac{SGWB} from cosmic string can be regarded as an ideal source for probing the expansion history of the Universe~\cite{Cui:2017ufi}. 
We also show the \acp{SNR} for the \ac{SGWB} of cosmic defects in \tab{tab:snr_cd}.

\begin{table}
	\begin{center}
		\caption{\acp{SNR} for the \ac{SGWB} from cosmic defects, assuming 1 year operation time.}\label{tab:snr_cd}
		\setlength{\tabcolsep}{0.5mm}
		\renewcommand\arraystretch{1.2}
		\begin{tabular}{|c|c|c|c|}
			\hline
		                          & string network & \multicolumn{2}{c|}{cosmic string} \\
		    \cline{2-4}
			                      & $N=3$          & $G\mu=10^{-11}$ &$G\mu=10^{-17}$   \\
			\hline 
			TQ                    & \multirow{2}*{$1.2\times10^{-1}$}&  \multirow{2}*{360}    &\multirow{2}*{$5.5\times10^{-1}$}   \\
			(null channel)        &                &                 &  \\
			\hline 
			TQ I+II               & \multirow{2}*{$5.0\times10^{-2}$}&  \multirow{2}*{160}    &\multirow{2}*{$2.3\times10^{-1}$}   \\
			(cross correlation)   &                &                 &  \\
			\hline
			TQ+LISA               & \multirow{2}*{$7.3\times10^{-2}$}&  \multirow{2}*{120}    &\multirow{2}*{$1.3\times10^{-1}$}   \\
			(cross correlation)   &                &                 &  \\
			\hline
		\end{tabular}
	\end{center}
\end{table}

\begin{table}
	\begin{center}
		\setlength{\tabcolsep}{3mm}
		\renewcommand\arraystretch{1.5}
		\caption{Order-of-magnitude range of \acp{SNR} for different types of \ac{SGWB} with TianQin/TianQin+LISA.}\label{tab:snr}
		\begin{tabular}{*{2}{|c}|}
			\hline
			Origins             & ${\rm SNR}$    \\
			\hline
			\ac{SBBH}+\ac{BNS}  & $\sim10$       \\
			\hline
			\ac{EDWD}           & $<10^{2}$       \\
			\hline 
			MBHB                & $<1$         \\
			\hline
			Primordial GW  & $\ll1$         \\
			\hline
			Phase transition    & $<10^{5}$      \\
			\hline
			Cosmic defects      & $<10^{3}$      \\
			\hline
			%%PBH                & $\sim10$  \\
			%\hline
		\end{tabular}
	\end{center}
\end{table}

%%%%%%%%%%%%%%%%%%%%
\section{Summary}\label{sec:summary}
We mean that in this work we study comprehensively the detection ability/potential of a number of different types of \ac{SGWB}.
\acp{SGWB} can be roughly divided into two categories, i.e., of either astrophysical or cosmological origin. 
We plot the expected signal strength for the \acp{SGWB} of these two categories in \figs{fig:astro} and \ref{fig:cos}, respectively. 

Due to the different formation mechanisms, the spectra of different sources have distinctive shapes. 
For example, the \ac{SGWB} from cosmic defects appears to be nearly scale-invariant in the TianQin band, the \acp{SBBH}, \acp{BNS}, and \acp{EDWD} \ac{SGWB} spectrum follows a power-law, while the \ac{SGWB} of the \ac{PT} has a parabolic shape.
As a comparison, we also plot the sensitivity curves in the form of \ac{PI} curves, assuming several configurations, including TianQin, TianQin I+II, and TianQin+LISA.  

Most of the astrophysical sources have existing observations to calibrate, and therefore the expected \ac{SGWB} include relatively small uncertainties. 
On the other hand, the \ac{SGWB} spectra generated from most cosmological sources have a large theoretical uncertainty in both the amplitude and the peak frequency, even the spectral shape can be quite exotic.
This is due to the fact that new physics models beyond the SM that drive inflation or \acp{PT} typically contain a large amount of model parameters which is not tightly constrained by the current experimental data. 
Such theoretical uncertainty also applies to other signals, like induced gravitational wave. Quantum fluctuations during inflation that reenter the horizon in the \ac{RD} era can collapse and form \acp{PBH}, and the scalar fluctuations can form \ac{SGWB}~\cite{Carr:1974nx,Saito:2008jc}. It is expected to peak at mHz, and if \acp{PBH} are responsible for all the dark matter, then the \ac{SGWB} can be detectable by TianQin~\cite{Cai:2018dig}.
Therefore, the present analysis is qualitatively reliable way to estimate the detectability of a GW signal. 
Notice that in order to probe new physics by means of GW detection, one has to isolate the SGWB produced from different sources~\cite{Biscoveanu:2020gds,Boileau:2020rpg}.

%===========disccusion of the foreground=============%
Among all discussed \acp{SGWB}, the strongest astrophysical source is expected to be Galactic \acp{DWD}. 
It can become comparable to the detector noise, and thus it is sometimes referred to as the foreground.
One can suppress the foreground with more observation time and by combining more detectors, and for the first time, we discussed a joint foreground with the TianQin-LISA detector network.

%===========discussion of the detection method of the background =============%
In order to detect \ac{SGWB}, one can use the cross correlation and null channel methods.
The channels within a triangular detector like TianQin, or LISA are colocated and share the same operation time.
The detection sensitivity of null channel method could benefit from the above factors.
However, the {\emph a priori} knowledge of the noise model needed for the null channel method makes it less appealing than the cross correlation method. 
Overall, the cross correlation method is more robust and reliable if a simultaneous observation of multiple detectors is possible, like a network of TianQin+LISA or TianQin I+II.

%======================\ac{SNR} estimate and comment==================%
Considering the uncertainties involved in the models, we presented a rough range of the expected \acp{SNR} for a number of \acp{SGWB} in \tab{tab:snr}, with an assumed operation time $T_{\rm op}=1\,{\rm yr}$. 
For example, the predicted \ac{SGWB} from \acp{SBBH} can differ greatly between the field binary evolution channel~\cite{Kowalska:2010qg}, the dynamical capture channel~\cite{Rodriguez:2016kxx}, and the active galactic nuclei disk channel~\cite{Antonini:2012ad}, since the predicted eccentricity distributions can be wildly different.
Therefore, the observation of astronomical sources will help us greatly advance our understanding of the properties of compact binary population. 
For the less certain cosmological sources, their observations will clearly lead to a long-awaited breakthrough beyond the SM. 
However, even a null observation can still help to place important upper limits on corresponding models and better guide theoretical developments.

Recently, the NANOGrav team reported an interesting observation from their 12.5-year data, where strong evidence points to the existence of a stochastic process that cannot be explained by noise.
However, due to the lack of a quadrupole spatial correlation, the NANOGrav team does not claim the detection of \ac{SGWB}~\cite{Arzoumanian:2020vkk}. 
This discovery adds credibility in SGWB detection, and we believe that the observations of nHz GW would certainly trigger a huge spike in the understanding of underlying physics behind the future \ac{SGWB} detections. 
On the other hand, the PPTA team find no evidence of the SGWB detection, 
but restrict the model parameters of first-order \ac{PT}, 
which can be employed in the dark and QCD \acp{PT}~\cite{Xue:2021gyq}. Furthermore, with the increasing sensitivity in the nHz frequency band, as well as the opening of high-frequency band of the GW window, we can expect to use multiband GW observation to better constrain our understanding of \acp{SGWB}~\cite{Colpi:2016fup}.

\begin{acknowledgments}
This work has been supported by the National Key Research and Development Program of China (No. 2020YFC2201400), and the Natural Science Foundation of China (Grants No. 11805286, 11690022). Y. J. is supported by the GuangDong Basic and Applied Basic Research Foundation (No. 2020A1515110150).
We would like to thank Valeriya Korol for providing the sample of Galaxy \ac{DWD}.
We thank the anonymous referee for valuable suggestions that improved the manuscript significantly.
We also thank Neil Cornish, Yi-Fan Wang, Shun-Jia Huang, Hai-Tian Wang, Xiang-Yu Lyu, Bo-Bing Ye, En-Kun Li and Fa-Peng Huang for helpful discussions.
\end{acknowledgments}
\appendix
%\section{The channel response}\label{appen:channelresponse}

%\bea
%\nn
%F_{\rm X}^{ab}(f,\hat{k})
%&=&\frac{1}{2}
%\bigg[\bigg(u_{3}^{a}u_{3}^{b}\mathcal{T}'(f,-\hat{u}_{3},\hat{k})-
%u_{2}^{a}u_{2}^{b}\mathcal{T}'(f,-\hat{u}_{2},\hat{k})\bigg)\\
%&+&\bigg(u_{2}^{a}u_{2}^{b}\mathcal{T}(f,-\hat{u}_{2},\hat{k})-
%u_{3}^{a}u_{3}^{b}\mathcal{T}(f,-\hat{u}_{3},\hat{k})\bigg)\bigg]
%\eea
%where
%\bea
%\mathcal{T}(f,\hat{u},\hat{k})=\frac{1}{2}
%\big[{\rm %sinc}\big(\frac{f}{2f_{*}}(1-\hat{k}\cdot\hat{u})\big)\exp\big(-i\frac{f}{2f_{*}}(3+\hat{k}\c%dot\hat{u})\big)
%+{\rm %sinc}\big(\frac{f}{2f_{*}}(1+\hat{k}\cdot\hat{u})\big)\exp\big(-i\frac{f}{2f_{*}}(1+\hat{k}\c%dot\hat{u})\big)\big]
%\eea
\begin{widetext}
\section{Coordinates for TianQin and LISA}\label{appen:coordinate_TL}

When calculating the \ac{ORF} of TianQin + LISA, we need to construct a uniform coordinate system for TianQin and LISA. In this coordinate system where the $x$ axis points in the direction of the vernal equinox.

The ecliptic coordinates for TianQin are comprised of two parts. The ecliptic coordinates for the center of the Earth are given by
\bea
\label{eq:co_tq1}
\nn
X(t)&=&R\cos(\alpha_{\rm TQ}(t))+\frac{1}{2}eR\big(\cos(2\alpha_{\rm TQ}(t))-3\big)+O(e^{2})\\
\nn
Y(t)&=&R\sin(\alpha_{\rm TQ}(t))+\frac{1}{2}eR\sin(2\alpha_{\rm TQ}(t))+O(e^{2})\\
Z(t)&=&0,
\eea
where $e=0.0167$, $R=1\,\,\rm AU$, $\alpha_{\rm TQ}(t)=2\pi f_{\rm m}t-\beta$, $f_{\rm m}=1/{\rm yr}$, and the longitude of perihelion $\beta=102.9^{\circ}$. The geocentric-ecliptic coordinates for TianQin are given by
\bea
\label{eq:co_tq2}
\nn
\tilde{x}_{n}(t)&=&R_{1}\big(\cos\phi_{\rm s}\sin\theta_{\rm s}\sin \alpha_{n}+\cos\alpha_{n}\sin\phi_{\rm s}\big)\\
\nn
\tilde{y}_{n}(t)&=&R_{1}\big(\sin\phi_{\rm s}\sin\theta_{\rm s}\sin\alpha_{n}-\cos\alpha_{n}\cos\phi_{\rm s}\big)\\
\tilde{z}_{n}(t)&=&-R_{1}\sin\alpha_{n}\cos\theta_{\rm s},
\eea
where $\alpha_{n}=2\pi f_{\rm sc}t+\kappa_n$ with $\kappa_n=\frac{2}{3}(n-1)\pi$, $R_{1}=1\times10^{5}\,\,\rm km$, $\theta_{\rm s}=-4.7^{\circ}$, $\phi_{\rm s}=120.5^{\circ}$, $f_{\rm sc}=1/(3.64\,\rm days)$, and $n=1,2,3$, denotes the three satellites of TianQin. 

By adding \eqs{eq:co_tq1} and (\ref{eq:co_tq2}), the ecliptic coordinates for the three satellites are obtained.

On the other hand, the ecliptic coordinates for LISA are written as
\bea
\nn
x_{n}'(t)&=&R\cos(\alpha_{\rm LISA}(t))+\frac{1}{2}eR\big(\cos(2\alpha_{\rm LISA}(t)-\kappa_n)\\
\nn
&-&3\cos\kappa_n\big)\\
\nn
y_{n}'(t)&=&R\sin(\alpha_{\rm LISA}(t))+\frac{1}{2}eR\big(\sin(2\alpha_{\rm LISA}(t)-\kappa_n)\\
\nn
&-&3\sin\kappa_n\big)\\
z_{n}'(t)&=&-\sqrt{3}eR\cos(\alpha_{n}'(t)-\kappa_n),
\eea
where $\alpha_{\rm LISA}(t)=2\pi f_{\rm m}t-\beta+\Delta \alpha$ with $\kappa_n=\frac{2}{3}(n-1)\pi$ with the relative phase $\Delta \alpha\simeq 20^{\circ}$, $e=0.0048$, and $R=1\,\,\rm AU$.
\end{widetext}

%%%%%%%%%%%%%%%%%%%%%%%%%%%%%%%%%%%%%%%%%%%%%%%%%%%%%%%%%%%%%%%%
\bibliographystyle{apsrev4-1}
%%%%%%%%%%%%%%%%%%%%%%%%%%%%%%%%%%%%%%%%%%%%%%%%%%%%%%%%%%%%%%%%
%%%%%%%%%%%%%%%%%%%%%%%%%%%%%%%%%%%%%%%%%%%%%%%%%%%%%%%%%%%%%%%%
\bibliography{TQ-SGWB}
\end{document}